\documentclass[prd,twocolumn,tightenlines,nofootinbib,superscriptaddress,10pt,aps]{revtex4-2}
\usepackage{bm,dcolumn,amsmath,graphicx,amsfonts,amssymb,hyperref,xcolor}

\definecolor{cite}{rgb}{0.,0.,0.9}
\hypersetup{colorlinks,linkcolor={cite},citecolor={cite},urlcolor={cite}}

\renewcommand{\v}[1]{\ensuremath{\boldsymbol{#1}}} 

\usepackage[normalem]{ulem}
\definecolor{newc}{rgb}{0.,0.6,0.4}

\renewcommand{\section}[1]{\vspace{0.85pt}\paragraph*{\textbf{\textit{\small{#1---}}}}}
\renewcommand{\subsection}[1]{\paragraph*{{\textit{\small{#1---}}}}}

\begin{document}

\title{\mbox{Ultralight Dark Matter Search with Space-Time Separated Atomic Clocks and Cavities}}

\date{\today}

\author{Melina~Filzinger}\altaffiliation{These authors contributed equally to this work.}
\affiliation{Physikalisch-Technische Bundesanstalt, Bundesallee 100, 38116 Braunschweig, Germany}

\author{Ashlee~R.~Caddell}\altaffiliation{These authors contributed equally to this work.}
\affiliation{School of Mathematics and Physics, The University of Queensland, Brisbane QLD 4072, Australia}

\author{Dhruv~Jani}
\affiliation{School of Mathematics and Physics, The University of Queensland, Brisbane QLD 4072, Australia}

\author{Martin~Steinel}
\affiliation{Physikalisch-Technische Bundesanstalt, Bundesallee 100, 38116 Braunschweig, Germany}

\author{Leonardo~Giani}
\affiliation{School of Mathematics and Physics, The University of Queensland, Brisbane QLD 4072, Australia}

\author{Nils~Huntemann}
\affiliation{Physikalisch-Technische Bundesanstalt, Bundesallee 100, 38116 Braunschweig, Germany}

%---

\author{Benjamin~M.~Roberts}\email[]{b.roberts@uq.edu.au}
\affiliation{School of Mathematics and Physics, The University of Queensland, Brisbane QLD 4072, Australia}

%-------------------------------------------------------------------------
\begin{abstract}\noindent
We devise and demonstrate a method to search for non-gravitational couplings of ultralight dark matter to standard model particles using space-time separated atomic clocks and cavity-stabilized lasers. 
By making use of space-time separated sensors, which probe different values of an oscillating dark matter field, we can search for couplings that cancel in typical local experiments. 
This provides sensitivity to both the temporal and spatial fluctuations of the field.
We demonstrate this method using existing data from a frequency comparison of lasers stabilized to two optical cavities connected via a 2220\,km fiber link [Schioppo {\sl et al.}, Nat.\ Commun.\ {\bf13}, 212 (2022)], and from the atomic clocks on board the Global Position System satellites.
Our analysis results in constraints on the coupling of scalar dark matter to electrons, $d_{m_e}$, for masses between $10^{-19}$\,eV$/c^2$ and $2\times10^{-15}$\,eV$/c^2$.
These are the first constraints on $d_{m_e}$ alone in this mass range.
\end{abstract}

\maketitle

%/*****************************************************
%/*****************************************************
%/*****************************************************

\noindent
Dark matter provides the best explanation for a wide range of astrophysical observations, however its nature remains a complete mystery~\cite{Bertone:2016nfn}. 
One intriguing and well-motivated class of models are those of ultralight scalar dark matter, with  particle masses well below the ${\rm eV}/c^2$ level~\cite{Kimball2022,Antypas:2022asj}.
To account for the observed galactic dark matter abundance, the particle number density in these models must be very high, meaning they may be treated as classical bosonic fields. 
In such regimes, it becomes possible and necessary to utilize the coherent wavelike nature of the fields in dark matter searches~\cite{Safronova:2017xyt}.

Interactions between dark scalar fields and standard model particles may lead to effective variations in certain fundamental constants~\cite{Olive:2001vz,*Olive:2007aj, Arvanitaki:2014faa, Stadnik:2015kia},
such as the fine-structure constant or Fermion masses.
Since atomic transition frequencies depend on these constants, atomic clocks have been used to search for their variation in the form of 
drifts~\cite{Uzan:2011,Rosenband:2008,Leefer:2013waa,Godun:2014naa,Huntemann:2014dya,McGrew2018,Lange:2020cul,Schwarz:2020upg,Filzinger:2023zrs},
transients~\cite{Derevianko:2013oaa,Wcislo:2016qng,Roberts:2017hla,Roberts:2018agv,Wcislo:2018ojh,Roberts:2019sfo}, 
and oscillations~\cite{VanTilburg:2015oza,Hees:2016gop,Wcislo:2018ojh,Antypas:2019qji,Savalle:2020vgz,Kennedy:2020bac,Beloy:2020tgz,Oswald:2021vtc,*Tretiak:2022ndx,Zhang:2022ewz,Kobayashi:2022vsf,Filzinger:2023zrs,Sherrill:2023zah}. 
At the same time, the lengths of solid bodies depend on the fine-structure constant and electron mass via the Bohr radius~\cite{Stadnik:2014tta}.
In stabilizing a laser to an optical cavity, length changes of the cavity spacer are translated into frequency changes of the laser, which can be measured with high precision in optical frequency comparisons. 
Consequently, ultrastable cavities may also act as probes for possible variations of fundamental constants~\cite{Stadnik:2014tta, Geraci:2018fax}. 
We use the term oscillator to indicate both atomic clocks and cavities. 

Searches for ultralight dark matter with such oscillators consist of comparing their resonant frequencies over time, and analyzing the resulting data for variations.
Typically, the measurements rely on comparing at least two co-located oscillators that feature different sensitivities to the investigated fundamental constants, since equal sensitivities would lead to signal cancellation in the measured frequency ratios.

In this Letter, we devise and demonstrate a method for using space-time separated sensors to search for ultralight dark matter via sub-Hz oscillations of fundamental constants. 
Due to their separation, the sensors probe different local values of the same oscillating dark-matter field. 
Therefore, the measured frequency ratio is sensitive to dark-matter couplings even if both oscillators feature the same sensitivity to the fundamental constants. 
This allows us to place independent constraints on oscillations of the electron mass, while other experiments probing this frequency range, which used co-located frequency comparisons, constrained oscillations in the fine-structure constant
or the electron-to-proton mass ratio.

To demonstrate the method, we conduct two analyses using existing data.
First, we search for oscillations in the frequency ratio between two spatially-separated lasers that are stabilized to optical cavities and connected by a 2220\,km fiber link~\cite{Schioppo:2022iqe}.
Then, we use timing data from the microwave atomic clocks on board the Global Positioning System (GPS) satellites.
The first data set is sensitive to the spatial fluctuations of the dark matter field due to its finite oscillation wavelength, while the second is sensitive to the temporal fluctuations due to the oscillation frequency.
We place constraints on the coupling of scalar dark matter to electrons, $d_{m_e}$,
in the mass range ($10^{-19}$ -- $2\times10^{-15}$)\,eV.
Our results are the only constraints on $d_{m_e}$ alone in this mass window.

Ultralight bosonic dark matter can be described by a scalar field of mass $m_\phi$ 
with classically oscillating non-relativistic solutions~\cite{Hui:2016ltb,Hees:2018fpg}
\begin{equation}\label{eq:phi(t)}
    \phi(t,\v{x}) = \phi_0 \cos\left(\omega t - \v{k}\cdot\v{x} \right),
\end{equation}
where $\omega$\,$\approx$\,$m_\phi$ is determined by the Compton frequency, and $k$\,$\approx $\,$m_\phi v$, with $v$\,$\sim$\,$10^{-3}$ the dark matter velocity (except where stated, we use units $\hbar$\,=\,$c$\,=\,1). 
Under the assumption that the field saturates the dark matter, its amplitude is
$\phi_0 = {\sqrt{2\rho_{\rm DM}}}/{m_\phi},$
where $\rho_{\rm DM}$ is the local dark matter density, typically taken to be $\approx$\,$0.4\,{\rm GeV}/{\rm cm^3}$~\cite{Freese:2012xd}.
The oscillations have a finite coherence time
$\tau_c=(2\pi/\omega)(c/\Delta v)^2$~\cite{Graham:2011qk}, due to the assumed dark matter velocity spread $\Delta v\approx 10^{-3} c$.

If the field has non-gravitational interactions with the standard model, signatures may become apparent in experiments.
We consider the linear couplings~\cite{Damour:2010rp, Hees:2018fpg}:
\begin{multline}\label{eq:Lint}
    \mathcal{L}_{\rm int} = \kappa\phi\Big[
     \frac{d_e}{4} F^{\mu\nu}F_{\mu\nu}
    - \frac{d_g}{2\widetilde g_3} G^{a\mu\nu}G^a_{\mu\nu}
        \\
        - \sum_{f=e,u,d} \left(d_{m_f}+\gamma_{q}d_g\right)  m_f\bar\psi_f\psi_f
    \Big],
\end{multline}
where $F$ is the electromagnetic tensor, 
$\psi_f$ are the Fermion (electron $e$ and light quark $u$,\,$d$) fields, 
$G^a$ is the gluon tensor, 
$\widetilde g_3$ is the effective QCD coupling (including running), 
$\gamma_q$ is the anomalous dimension giving the mass running of QCD-coupled Fermions, and the $d$ factors are dimensionless coupling constants.
The factor 
$\kappa$\,$=$\,$\sqrt{4\pi}/{M_{\rm Pl}}$
is introduced to make the $d$ couplings dimensionless
($M_{\rm Pl}$\,$\approx$\,$1.2\times10^{19}$\,GeV is the Planck mass). 

The above interactions can be parameterized as effective variations in the fundamental constants~\cite{Damour:2010rp},
\begin{align}
    X(\phi) &= X\left(1 + d_X \, \kappa\phi\right),
\end{align}
where 
$X$ may be
$\alpha$, the fine structure constant (with $d_e$), $\Lambda_{g}$ the QCD mass scale (with $d_g$), or $m_f$ the Fermion masses.
These lead to variations in oscillator frequencies,
\begin{equation}
    \frac{\delta\nu}{\nu} 
    = \sum_X d_{X} K_X \, \kappa\phi
    \equiv d_{\rm eff}  \, \kappa\phi,
\end{equation}
where 
$K_X$ is the sensitivity of the particular frequency to variation in the fundamental constant, and we defined the effective coupling $d_{\rm eff}$ for convenience. 
The same phenomenology can arise for scalar and pseudoscalar particles (e.g., axions) with quadratic interactions~\cite{Kim:2023pvt,Banerjee:2022sqg,Flambaum:2023bnw,Beadle:2023flm}.

For optical atomic transitions, the coupling is~\cite{Savalle:2019jsb}:
\begin{equation}\label{eq:deff-optical}
    d^{\rm Opt}_{\rm eff} = (2 + K_{\rm rel})\,d_e + d_{m_e},
\end{equation}
where $K_{\rm rel}$ is a relativistic correction that depends on the specific transition~\cite{Dzuba:1999zz}.
For the hyperfine transitions employed in microwave clocks, it is~\cite{Flambaum:2004tm,*Flambaum:2006ip,Hees:2018fpg}
\begin{equation}\label{eq:deff-mw}
    d^{\rm mw}_{\rm eff} = (4 + K_{\rm rel})\,d_e + 2d_{m_e} - d_{g} + \kappa_q(d_{m_q} - d_{g}),
\end{equation}
where the dependence on 
$m_q=(m_u+m_d)/2$
stems from the dependence on the nuclear magnetic moment~\cite{Flambaum:2004tm,*Flambaum:2006ip}.
For both cases, there can be small additional contributions from the quark/gluon sectors, which enter via the nuclear radius or mass~\cite{Banerjee:2023bjc,Flambaum:2023drb}; these are not important for the current search.
The frequency of a laser stabilized to an optical cavity features the effective coupling
\begin{equation}\label{eq:deff-cavity}
    d^{\rm Cav}_{\rm eff} 
    = d_e + d_{m_e},
\end{equation}
due to field-induced changes in the cavity length.
Here, relativistic corrections depend on the cavity material, and are typically very small~\cite{Stadnik:2014tta}.

In typical local experiments, the frequency ratio of a pair of co-located oscillators $A$ and $B$ is observed.
Only if the oscillators have {\em different} effective field couplings, the ratio will contain a field-induced contribution
\begin{align}\label{eq:dnu-colocated}
    \frac{\delta(\nu^A/\nu^B)}{\nu^A/\nu^B}
        &= \kappa\phi_0\,\Delta d\cos(\omega t),
\end{align}
where $\Delta d$\,$=$\,$d^A_{\rm eff}$\,$-$\,$d^B_{\rm eff}$. 

Indeed, for co-located comparisons at low frequency, terms containing $d_{m_e}$ never appear alone, but rather in some relative combination~\cite{Kozlov:2018qid,Savalle:2020vgz} (see Refs.~\cite{Geraci:2018fax,Antypas:2019yvv} for discussions on high-frequency comparisons).
Hyperfine transitions depend on the electron-to-proton mass ratio $m_e/m_p$, and consequently comparing an optical to a microwave transition leads to effective couplings including $|d_{m_e}$\,$-$\,$d_g|$~\cite{Flambaum:2004tm,*Flambaum:2006ip}. 
This is important since in the simplest models $d_{m_e}$ and $d_g$ may be similar or equal~\cite{Damour:2010rp}.

A pair of space-time separated oscillators, on the other hand, will experience a different local value of the scalar field due to its oscillations.
In this case, the frequency ratio for a pair of oscillators with identical effective couplings will gain the field-induced shift
\begin{align}
    \frac{\delta(\nu^A/\nu^B)}{\nu^A/\nu^B}
        &= \kappa\phi_0 \, d_{\rm eff}\left[\cos(\omega t) - \cos(\omega t - \delta)\right]
        \notag\\
        \label{eq:signal-general}
        &\approx  \kappa\phi_0 \, d_{\rm eff} \sin(\omega t) \, \delta,
\end{align}
where $|\delta|$\,$\ll$\,1 is the scalar field phase difference between the two oscillators. 

Two terms contribute to the phase difference $\delta$:\ the time delay $\Delta t$ due to the signal propagation between the two oscillators, and a term due to the spatial separation
\begin{align}\label{eq:delta}
    \delta &= \omega \Delta t - \v{k}\cdot\Delta\v{x},
\end{align}
where 
$\Delta\v{x} = D\,\v{n}$, with
 ${D}$ the linear distance between the oscillators, and $\v{n}$ the separation unit vector.
In the considered mass range $m_\phi$\,$\lesssim $\,$10^{-14}\,{\rm eV}$, we have $\lambda_\phi$\,$=$\,$2\pi/k\gtrsim 10^8\,{\rm km}$\,$\gg$\,$D$, so the field can be considered coherent across all oscillators.
Notice that
${k D}/({\omega\Delta t}) \approx v/c\approx 10^{-3}$.
Therefore, the contribution of the spatial phase difference can be neglected, unless the contribution of the temporal term is experimentally suppressed.

For a dark matter interpretation, $\phi_0$\,$\propto$\,$m_\phi^{-1}$, while $\omega$\,$=$\,$m_\phi$.
At the same time, both terms in the phase shift \eqref{eq:delta} are linear in $m_\phi$.
Therefore, the signal amplitude~\eqref{eq:signal-general} would be independent of the dark matter mass. 
This is in contrast to most dark matter searches, where the signal amplitude would scale inversely with the mass. 

The signal~\eqref{eq:signal-general} has a linear scaling with the effective distance ($D$ or $\Delta t / c$) between the oscillators, 
which offers a method to enhance the sensitivity with spatially separated sensors. 
Crucially, the dependence on distance may also act as a key dark matter signature:\ 
A network of comparisons with multiple distances could be used to exclude spurious terrestrial sources for the oscillations.

%======================================================

The network of optical fiber-linked atomic clocks and cavities located at PTB in Germany, LNE-SYRTE in France, INRIM in Italy, and NPL in the United Kingdom~\cite{Lopez:2012, Raupach:2015, Chiodo:2015oma, Lisdat2016, Koke:2019bwo, ITALY2022} is particularly suitable for precision frequency comparisons over large distances. 
The fiber links allow state-of-the-art comparisons of optical frequencies over $\sim$\,$10^3$\,km distances with $\lesssim$\,$10^{-17}$ precision~\cite{Lisdat2016,Schioppo:2022iqe}.
The network has been employed previously for tests of fundamental physics~\cite{Delva:2017lie}, and to search for transient variations in fundamental constants and dark matter~\cite{Roberts:2019sfo}.

As an initial search demonstrating  our method, we analyze publicly available data from a pair of ultra-stable lasers located at NPL and PTB~\cite{Schioppo:2022iqe,SchioppoZenodo:2021}.
The frequencies of the cavity-stabilized lasers are compared via the optical fiber links, and have been measured to achieve a fractional frequency instability as low as $7\times10^{-17}$~\cite{Schioppo:2022iqe}. 

The fiber-link distance $L$ between the cavities is 2220\,km~\cite{Schioppo:2022iqe}, corresponding to a delay time of 
$\approx$\,$10\,{\rm ms}$.
However, due to the active noise cancellation employed, the time delay contribution leads to no observable effects in the frequency comparisons.
The noise cancellation works by sending a frequency signal from one oscillator to the other, reflecting it, and comparing to its source~\cite{Lisdat2016}.
Any phase shift between the sent and received signals is assumed to be noise introduced on the fiber link and actively suppressed.
Since the test signal traverses the fiber length twice, half the observed phase shift is subtracted from the signal.
In the presence of the dark matter field, the test signal would gain a dark-matter--induced shift in comparison to its source~[Eq.~\eqref{eq:signal-general}] due to the time delay between the sent and received signals with $\delta = 2\omega\Delta t$ (where $\Delta t = nL/c$ is the one-way time delay with the fiber's refractive index $n$).
The noise-correction signal would thus contain a component exactly equal to the time-delay contribution to the dark matter signal.
As such, any signal caused by a time delay would be interpreted as noise, and removed.
Since the considered dark-matter oscillation period is much larger than the fiber delay time,
any signal originating from dark-matter--induced modifications to the length and refractive index of the fiber~\cite{Savalle:2020vgz} are also removed by the noise cancellation.

%%%%%%%%%%%%%%%%%%%%%%%%%%%%%%%%%%%%%%%%%%%%%%%%%%%%%%%%%%%

On the other hand, the signal coming from the spatial separation is not removed.
This provides a unique opportunity to probe the spatial fluctuations of scalar dark matter, with induced signal
\begin{equation}\label{eq:signal-spatial}
    \frac{\delta(\nu^A/\nu^B)}{\nu^A/\nu^B} = 
    \kappa\phi_0\,
    (d_{m_e} + d_e)\,
    \frac{\omega D \, \v{n}\cdot\v{v}}{c^2}\,\sin(\omega t).
\end{equation}

To determine the signal strength, we average Eq.~\eqref{eq:signal-spatial}
over the dark matter velocity distribution, $f(\v{v})$.
In the galactic rest frame, the average would be zero.
In the Earth frame, however, it is non-zero due to its motion through the galaxy with speed $v_{\odot}$\,$\approx$\,$220\,{\rm km}\,{\rm s}^{-1}$ (roughly towards the Cygnus constellation). We have
\begin{equation}
    \int {\rm d}^3 \v{v} \, f(\v{v})  \, \v{v}\cdot \v{n}
    = \v{v}_{\odot}\cdot\v{n}
    = {v_{\odot}}\cos\gamma(t),
\end{equation}
where 
$\gamma$ is the angle between the laboratory separation and the direction of Earth's galactic motion, which we compute using Ref~\cite{Hirsch:2016}.
This exhibits a strong sidereal modulation due to Earth's rotation.

In general, this modulation would present a key dark matter observable.
For the present data set, however, there is less than one full day of data~\cite{SchioppoZenodo:2021}.
As such, we do not search for the daily modulation, but rather average the expected signal amplitude over the observation period, finding
$ |\langle{\cos\gamma}\rangle_{\rm T_{\rm obs}}|\approx 0.41$~\cite{GitHubGPS}.

%%%%%%%%%%%%%%%%%%%%%%%%%%%%%%%%%%%%%%%%%%%%%%%%%%%%%%%%%%%

\begin{figure}
\includegraphics[width=\linewidth]{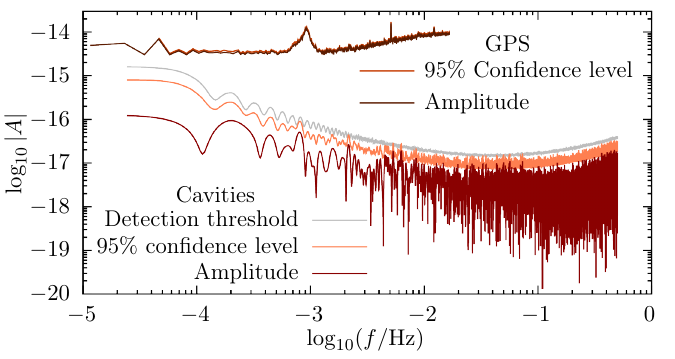}
\caption{Amplitude spectrum $A$ of the NPL/PTB cavity-cavity comparison, and GPS microwave clock data, showing the upper 95\% confidence levels and detection threshold.}
\label{fig:amplitude}
\end{figure}

We search for oscillations in the form of Eq.~\eqref{eq:signal-spatial} in the data,
following the method presented in Ref.~\cite{Filzinger:2023zrs,Hees:2016gop}. 
The best-fit amplitudes of sinusoidal oscillations are extracted using the Lomb-Scargle formalism~\cite{VanderPlas_2018}; the resulting  spectrum is shown in Fig.~\ref{fig:amplitude}.
The 95\% confidence level and the detection threshold are determined using Monte-Carlo sampling, implementing the full noise model as detailed in Table 1 of Ref.~\cite{SchioppoZenodo:2021}.
The confidence levels are obtained by adding a randomly generated offset to each data point according to the noise model 1000 times, generating datasets with independent noise realizations. 
The detection threshold is similarly obtained, but is based on pure noise, and takes into account the look-elsewhere effect~\cite{VanderPlas_2018,Hees:2016gop}. 
The high-frequency bound of our analysis is 0.5\,Hz, limited by the 1\,Hz sampling rate, while the low-frequency bound is determined by the total measurement duration of $4.1\times10^4$\,s. 
For frequencies $f$\,$=$\,$\omega/(2\pi)$\,$<$\,${\rm 1\, Hz}$, the dark matter coherence time is $\tau_c$\,$>$\,$10^6$\,${\rm s}$, 
well above the total measurement duration.

\begin{figure*}
\includegraphics[width=0.99\linewidth]{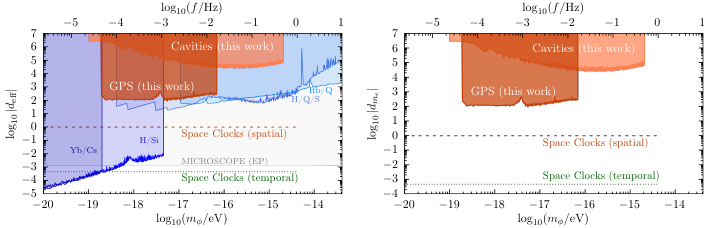}
\caption{Constraints on effective couplings $d_{\text{eff}}$ involving $d_{m_e}$ (95\% confidence level) as a function of the dark matter mass $m_\phi$ and corresponding oscillation frequency $f=\omega/2\pi$ (left). 
Shown in blue are existing constraints on $|d_{m_e}-d_g|$ from local clock and oscillator comparisons (Yb/Cs~\cite{Kobayashi:2022vsf},
H/Si~\cite{Kennedy:2020bac},
H/Quartz/Sapphire~\cite{Campbell:2020fvq}, and
Rb/Quartz~\cite{Zhang:2022ewz}; 
the Ref.~\cite{Campbell:2020fvq} constraint has been rescaled 
to account for the stochasticity~\cite{Centers:2019dyn}, which is already included in the other constraints).
The gray band shows constraints on $|d_{m_e}-d_g|$ from equivalence principle tests (MICROSCOPE~\cite{Touboul:2017grn,*Berge:2017ovy,Hees:2018fpg}).
The constraints from this work, shown in orange, are the only constraints on $d_{m_e}$ alone in this mass range (right).
}
\label{fig:limits}
\end{figure*}

We find no statistically significant oscillations, which would appear as amplitude peaks above the detection threshold, and therefore place constraints on the magnitude of $|d_e + d_{m_e}|$. 
By combining with existing constraints on $|d_e|$~\cite{Beloy:2020tgz, Zhang:2022ewz, Kobayashi:2022vsf, Filzinger:2023zrs, Sherrill:2023zah}, which are at least an order of magnitude tighter, we place the first constraints on $|d_{m_e}|$ alone in this mass range, which we present in Fig.~\ref{fig:limits}. 
The previous experiments in this range probe combinations including $|d_{m_e}$\,$-$\,$d_g|$.
In the analysis, we account for the stochastic nature of the scalar fields~\cite{Derevianko:2016vpm,Centers:2019dyn}:\
For measurement durations below the coherence time, only the coherent part of the field density, $\phi_{\rm c} = \xi\phi_0$, is observable, where $\xi\approx1/3$ at the 95\% confidence level~\cite{Centers:2019dyn}.

As a complementary example, we also consider data from the Rb microwave clocks on board the GPS satellites, as compared to an Earth-based hydrogen maser.
The GPS data have previously been used for dark matter and cosmic field searches~\cite{Derevianko:2013oaa,Roberts:2017hla,Li:2023qpx}.
The GPS comparisons are made with a one-way microwave link, and the type of noise cancellation as employed in the fiber network is not possible.
Therefore, this search is sensitive to the phase difference caused by the signal time delay:
\begin{equation}\label{eq:signal}
    \frac{\delta(\nu^A/\nu^B)}{\nu^A/\nu^B} = 
    \kappa\phi_0\,d^{\rm mw}_{\rm eff}\,
    \frac{\omega D}{c}\,\sin(\omega t).
\end{equation}

The GPS constellation consists of approximately 30 satellites in orbit at a radius of $D\approx26\,000\,{\rm km}$.
Signals driven by an onboard Rb microwave atomic clock are broadcast from each of the satellites.
The time differences between the satellite clocks and a ground-based hydrogen maser are determined, and made publicly available~\cite{MurphyJPL2015,*JPLigsac}.
From these, we derive the fractional frequency differences averaged over the 30\,s sampling period~\cite{Roberts:2017hla}.

Relativistic Doppler shift effects cause periodic variations in the clock timings that oscillate at the $\sim$\,12\,${\rm hr}$ orbital period.
These are modeled and removed~\cite{Blewitt2015307}. 
Due to the imperfect nature of the modeling, residual effects are present in the data, and consequently, we do not search for oscillations at or below this frequency.

Since the scalar field is coherent across the network, and the time delay between each of the satellites and the Earth-based receiver station is roughly the same, the dark matter signal should be approximately equal for each GPS satellite clock.
Therefore, for each day of data, we form a weighted average of the frequency comparisons for each of the Rb satellite clocks in the network,
compared to the common ground-based H-maser.
We then form the amplitude spectrum for each day of data.
Since we only use clocks for which there are no missing data points, we use the standard Fourier method.
% 6242 complete "clock days", with 1246 discarded. On average, 26 clocks per day with no missing data
We repeat the process for the most recent 34 weeks of data 
% (between September 17, 2023 and May 11, 2024), 
(up to May 11, 2024), 
resulting in 238 spectra.
We take the average as the best-fit amplitude spectrum, and use the spread in amplitudes at each frequency to determine the maximum spectrum at the 95\% confidence level.
Due to the large number of spectra, this is very close to the average, as shown in Fig.~\ref{fig:amplitude}.
We have made our code for this analysis public~\cite{GitHubGPS}.

There are several large peaks in the GPS amplitude spectrum, which are likely caused by the satellite operation and signal processing. Without a full noise model for these clocks, we cannot define a detection threshold.
From the analysis, we place constraints on the combination of couplings in Eq.~\eqref{eq:deff-mw}.
By combining with existing constraints on 
$|d_e|$~\cite{Beloy:2020tgz, Zhang:2022ewz, Kobayashi:2022vsf, Filzinger:2023zrs, Sherrill:2023zah}, 
$|d_{m_e}-d_g|$
~\cite{Kobayashi:2022vsf,Kennedy:2020bac,Campbell:2020fvq,Zhang:2022ewz},
and $|d_{m_q} - d_g|$~\cite{Hees:2016gop,Kobayashi:2022vsf},
we constrain $d_{m_e}$ alone.
The 95\% C.L. constraints are shown in Fig.~\ref{fig:limits}.

In Fig.~\ref{fig:limits}, we also present projections of future sensitivities.
The ``Space Clocks'' projections are based on a future network of clocks showing white frequency noise with an amplitude of $10^{-18}$, and a separation equal to the GPS diameter~\cite{Delva:2017znr,AEDGE:2019nxb,Derevianko:2021kye,Schkolnik:2022utn,Tsai:2021lly}.
We show this both for the spatial term [Eq.~\eqref{eq:signal-spatial}], and the temporal term [Eq.~\eqref{eq:signal}], which would apply only if the noise cancellation does not remove the time-delay contribution signal.

In conclusion, we present a method to search for ultralight dark matter with space-time separated atomic clocks and cavities. 
We demonstrate this method by searching for oscillations with periods ranging from 2\,s to $10^5$\,s using frequency ratio measurements of lasers stabilized to optical cavities, and the microwave clocks on board the GPS satellites. 
We placed constraints on the coupling of scalar dark matter to electrons in the mass range ($10^{-19}$ -- $2\times10^{-15}$)\,eV.
These are the only constraints on $d_{m_e}$ alone at low frequencies $<1$\,Hz.

Searching for ultralight dark matter with space-time separated oscillators  uniquely enables probing the spatial fluctuations of scalar dark matter, and features key signatures to distinguish a dark matter signal from a spurious signal in case of a detection, namely the scaling with the effective distance, and the sidereal modulation.
The method also circumvents the need for the oscillators to feature a differential sensitivity to the considered dark matter coupling. 
Without the need to optimize for large differential sensitivities, future comparisons between space-time separated oscillators are free to focus on utilizing the full potential of the available sensor types to investigate different frequency ranges. 
This is particularly relevant for future measurement campaigns involving long fiber links, and for  future fundamental physics investigations in space.

%======================================================
%\begin{acknowledgments}
We thank Ekkehard Peik for helpful suggestions on the manuscript,
and Zachary Stevens-Hough for carefully reading the manuscript.
This work was supported by Australian Research Council (ARC) DECRA Fellowship DE210101026.
BMR acknowledges the support from the UQ Fellowship of the Big Questions Institute.
LG acknowledges support via ARC Fellowship FL180100168.
This research was supported by the Munich Institute for Astro-, Particle and BioPhysics (MIAPbP) which is funded by the Deutsche Forschungsgemeinschaft (DFG, German Research Foundation) under Germany's Excellence Strategy -- EXC-2094 -- 390783311.
This work was supported by the Max Planck–RIKEN–PTB Center for Time, Constants and Fundamental Symmetries, by the
Deutsche Forschungsgemeinschaft (DFG, German Research Foundation) under SFB 1227 DQ-mat -- Project-ID 274200144 -- within project B02, and the project 22IEM01 TOCK. 
The project (22IEM01 TOCK) has received funding from the European Partnership on Metrology, co-financed from the European Union’s Horizon Europe Research and Innovation Programme and by the Participating States.
The figures in this Letter were produced with the aid of Ref.~\cite{AxionLimits}.
%\end{acknowledgments}

%======================================================
\bibliography{dme}

%apsrev4-2.bst 2019-01-14 (MD) hand-edited version of apsrev4-1.bst
%Control: key (0)
%Control: author (8) initials jnrlst
%Control: editor formatted (1) identically to author
%Control: production of article title (0) allowed
%Control: page (0) single
%Control: year (1) truncated
%Control: production of eprint (0) enabled
\begin{thebibliography}{80}%
\makeatletter
\providecommand \@ifxundefined [1]{%
 \@ifx{#1\undefined}
}%
\providecommand \@ifnum [1]{%
 \ifnum #1\expandafter \@firstoftwo
 \else \expandafter \@secondoftwo
 \fi
}%
\providecommand \@ifx [1]{%
 \ifx #1\expandafter \@firstoftwo
 \else \expandafter \@secondoftwo
 \fi
}%
\providecommand \natexlab [1]{#1}%
\providecommand \enquote  [1]{``#1''}%
\providecommand \bibnamefont  [1]{#1}%
\providecommand \bibfnamefont [1]{#1}%
\providecommand \citenamefont [1]{#1}%
\providecommand \href@noop [0]{\@secondoftwo}%
\providecommand \href [0]{\begingroup \@sanitize@url \@href}%
\providecommand \@href[1]{\@@startlink{#1}\@@href}%
\providecommand \@@href[1]{\endgroup#1\@@endlink}%
\providecommand \@sanitize@url [0]{\catcode `\\12\catcode `\$12\catcode
  `\&12\catcode `\#12\catcode `\^12\catcode `\_12\catcode `\%12\relax}%
\providecommand \@@startlink[1]{}%
\providecommand \@@endlink[0]{}%
\providecommand \url  [0]{\begingroup\@sanitize@url \@url }%
\providecommand \@url [1]{\endgroup\@href {#1}{\urlprefix }}%
\providecommand \urlprefix  [0]{URL }%
\providecommand \Eprint [0]{\href }%
\providecommand \doibase [0]{https://doi.org/}%
\providecommand \selectlanguage [0]{\@gobble}%
\providecommand \bibinfo  [0]{\@secondoftwo}%
\providecommand \bibfield  [0]{\@secondoftwo}%
\providecommand \translation [1]{[#1]}%
\providecommand \BibitemOpen [0]{}%
\providecommand \bibitemStop [0]{}%
\providecommand \bibitemNoStop [0]{.\EOS\space}%
\providecommand \EOS [0]{\spacefactor3000\relax}%
\providecommand \BibitemShut  [1]{\csname bibitem#1\endcsname}%
\let\auto@bib@innerbib\@empty
%</preamble>
\bibitem [{\citenamefont {Bertone}\ and\ \citenamefont
  {Hooper}(2018)}]{Bertone:2016nfn}%
  \BibitemOpen
  \bibfield  {author} {\bibinfo {author} {\bibfnamefont {G.}~\bibnamefont
  {Bertone}}\ and\ \bibinfo {author} {\bibfnamefont {D.}~\bibnamefont
  {Hooper}},\ }\bibfield  {title} {\bibinfo {title} {A {{History}} of {{Dark
  Matter}}},\ }\href {https://doi.org/10.1103/RevModPhys.90.045002} {\bibfield
  {journal} {\bibinfo  {journal} {Rev. Mod. Phys.}\ }\textbf {\bibinfo {volume}
  {90}},\ \bibinfo {pages} {45002} (\bibinfo {year} {2018})},\ \Eprint
  {https://arxiv.org/abs/1605.04909} {arXiv:1605.04909} \BibitemShut {NoStop}%
\bibitem [{\citenamefont {Jackson~Kimball}\ and\ \citenamefont {van
  Bibber}(2023)}]{Kimball2022}%
  \BibitemOpen
  \bibinfo {editor} {\bibfnamefont {D.~F.}\ \bibnamefont {Jackson~Kimball}}\
  and\ \bibinfo {editor} {\bibfnamefont {K.}~\bibnamefont {van Bibber}},\
  eds.,\ \href {https://link.springer.com/book/10.1007/978-3-030-95852-7}
  {\emph {\bibinfo {title} {{The Search for Ultralight Bosonic Dark Matter}}}}\
  (\bibinfo {year} {2023})\BibitemShut {NoStop}%
\bibitem [{\citenamefont {Antypas}\ \emph {et~al.}(2022)\citenamefont {Antypas}
  \emph {et~al.}}]{Antypas:2022asj}%
  \BibitemOpen
  \bibfield  {author} {\bibinfo {author} {\bibfnamefont {D.}~\bibnamefont
  {Antypas}} \emph {et~al.},\ }\bibfield  {title} {\bibinfo {title} {New
  {{Horizons}}:\ {{Scalar}} and {{Vector Ultralight Dark Matter}}},\ }\href
  {http://arxiv.org/abs/2203.14915} {\bibfield  {journal} {\bibinfo  {journal}
  {Snowmass 2021 White Paper}\ } (\bibinfo {year} {2022})},\ \Eprint
  {https://arxiv.org/abs/2203.14915} {arXiv:2203.14915} \BibitemShut {NoStop}%
\bibitem [{\citenamefont {Safronova}\ \emph {et~al.}(2018)\citenamefont
  {Safronova}, \citenamefont {Budker}, \citenamefont {Demille}, \citenamefont
  {Jackson~Kimball}, \citenamefont {Derevianko},\ and\ \citenamefont
  {Clark}}]{Safronova:2017xyt}%
  \BibitemOpen
  \bibfield  {author} {\bibinfo {author} {\bibfnamefont {M.~S.}\ \bibnamefont
  {Safronova}}, \bibinfo {author} {\bibfnamefont {D.}~\bibnamefont {Budker}},
  \bibinfo {author} {\bibfnamefont {D.}~\bibnamefont {Demille}}, \bibinfo
  {author} {\bibfnamefont {D.~F.}\ \bibnamefont {Jackson~Kimball}}, \bibinfo
  {author} {\bibfnamefont {A.}~\bibnamefont {Derevianko}},\ and\ \bibinfo
  {author} {\bibfnamefont {C.~W.}\ \bibnamefont {Clark}},\ }\bibfield  {title}
  {\bibinfo {title} {Search for new physics with atoms and molecules},\ }\href
  {https://doi.org/10.1103/RevModPhys.90.025008} {\bibfield  {journal}
  {\bibinfo  {journal} {Rev. Mod. Phys.}\ }\textbf {\bibinfo {volume} {90}},\
  \bibinfo {pages} {025008} (\bibinfo {year} {2018})},\ \Eprint
  {https://arxiv.org/abs/1710.01833} {arXiv:1710.01833} \BibitemShut {NoStop}%
\bibitem [{\citenamefont {Olive}\ and\ \citenamefont
  {Pospelov}(2002)}]{Olive:2001vz}%
  \BibitemOpen
  \bibfield  {author} {\bibinfo {author} {\bibfnamefont {K.~A.}\ \bibnamefont
  {Olive}}\ and\ \bibinfo {author} {\bibfnamefont {M.}~\bibnamefont
  {Pospelov}},\ }\bibfield  {title} {\bibinfo {title} {Evolution of the fine
  structure constant driven by dark matter and the cosmological constant},\
  }\href {https://doi.org/10.1103/PhysRevD.65.085044} {\bibfield  {journal}
  {\bibinfo  {journal} {Phys. Rev. D}\ }\textbf {\bibinfo {volume} {65}},\
  \bibinfo {pages} {085044} (\bibinfo {year} {2002})},\ \Eprint
  {https://arxiv.org/abs/hep-ph/0110377} {arXiv:hep-ph/0110377} \BibitemShut
  {NoStop}%
\bibitem [{\citenamefont {Olive}\ and\ \citenamefont
  {Pospelov}(2008)}]{Olive:2007aj}%
  \BibitemOpen
  \bibfield  {author} {\bibinfo {author} {\bibfnamefont {K.~A.}\ \bibnamefont
  {Olive}}\ and\ \bibinfo {author} {\bibfnamefont {M.}~\bibnamefont
  {Pospelov}},\ }\bibfield  {title} {\bibinfo {title} {Environmental dependence
  of masses and coupling constants},\ }\href
  {https://doi.org/10.1103/PhysRevD.77.043524} {\bibfield  {journal} {\bibinfo
  {journal} {Phys. Rev. D}\ }\textbf {\bibinfo {volume} {77}},\ \bibinfo
  {pages} {043524} (\bibinfo {year} {2008})},\ \Eprint
  {https://arxiv.org/abs/0709.3825} {arXiv:0709.3825} \BibitemShut {NoStop}%
\bibitem [{\citenamefont {Arvanitaki}\ \emph {et~al.}(2015)\citenamefont
  {Arvanitaki}, \citenamefont {Huang},\ and\ \citenamefont
  {Van~Tilburg}}]{Arvanitaki:2014faa}%
  \BibitemOpen
  \bibfield  {author} {\bibinfo {author} {\bibfnamefont {A.}~\bibnamefont
  {Arvanitaki}}, \bibinfo {author} {\bibfnamefont {J.}~\bibnamefont {Huang}},\
  and\ \bibinfo {author} {\bibfnamefont {K.}~\bibnamefont {Van~Tilburg}},\
  }\bibfield  {title} {\bibinfo {title} {Searching for dilaton dark matter with
  atomic clocks},\ }\href {https://doi.org/10.1103/PhysRevD.91.015015}
  {\bibfield  {journal} {\bibinfo  {journal} {Phys. Rev. D}\ }\textbf {\bibinfo
  {volume} {91}},\ \bibinfo {pages} {015015} (\bibinfo {year} {2015})},\
  \Eprint {https://arxiv.org/abs/1405.2925} {arXiv:1405.2925} \BibitemShut
  {NoStop}%
\bibitem [{\citenamefont {Stadnik}\ and\ \citenamefont
  {Flambaum}(2015{\natexlab{a}})}]{Stadnik:2015kia}%
  \BibitemOpen
  \bibfield  {author} {\bibinfo {author} {\bibfnamefont {Y.~V.}\ \bibnamefont
  {Stadnik}}\ and\ \bibinfo {author} {\bibfnamefont {V.~V.}\ \bibnamefont
  {Flambaum}},\ }\bibfield  {title} {\bibinfo {title} {{Can Dark Matter Induce
  Cosmological Evolution of the Fundamental Constants of Nature?}},\ }\href
  {https://doi.org/10.1103/PhysRevLett.115.201301} {\bibfield  {journal}
  {\bibinfo  {journal} {Phys. Rev. Lett.}\ }\textbf {\bibinfo {volume} {115}},\
  \bibinfo {pages} {201301} (\bibinfo {year} {2015}{\natexlab{a}})},\ \Eprint
  {https://arxiv.org/abs/1503.08540} {arXiv:1503.08540} \BibitemShut {NoStop}%
\bibitem [{\citenamefont {Uzan}(2011)}]{Uzan:2011}%
  \BibitemOpen
  \bibfield  {author} {\bibinfo {author} {\bibfnamefont {J.-P.}\ \bibnamefont
  {Uzan}},\ }\bibfield  {title} {\bibinfo {title} {Varying constants,
  gravitation and cosmology},\ }\href {https://doi.org/10.12942/lrr-2011-2}
  {\bibfield  {journal} {\bibinfo  {journal} {Living Rev. Relativ.}\ }\textbf
  {\bibinfo {volume} {14}},\ \bibinfo {pages} {2} (\bibinfo {year} {2011})},\
  \Eprint {https://arxiv.org/abs/1009.5514} {arXiv:1009.5514} \BibitemShut
  {NoStop}%
\bibitem [{\citenamefont {Rosenband}\ \emph {et~al.}(2008)\citenamefont
  {Rosenband} \emph {et~al.}}]{Rosenband:2008}%
  \BibitemOpen
  \bibfield  {author} {\bibinfo {author} {\bibfnamefont {T.}~\bibnamefont
  {Rosenband}} \emph {et~al.},\ }\bibfield  {title} {\bibinfo {title}
  {{Frequency Ratio of Al${}^+$ and Hg${}^+$ Single-Ion Optical Clocks;
  Metrology at the 17th Decimal Place}},\ }\href
  {https://doi.org/10.1126/science.1154622} {\bibfield  {journal} {\bibinfo
  {journal} {Science}\ }\textbf {\bibinfo {volume} {319}},\ \bibinfo {pages}
  {1808} (\bibinfo {year} {2008})}\BibitemShut {NoStop}%
\bibitem [{\citenamefont {Leefer}\ \emph {et~al.}(2013)\citenamefont {Leefer},
  \citenamefont {Weber}, \citenamefont {Cing{\"o}z}, \citenamefont
  {Torgerson},\ and\ \citenamefont {Budker}}]{Leefer:2013waa}%
  \BibitemOpen
  \bibfield  {author} {\bibinfo {author} {\bibfnamefont {N.}~\bibnamefont
  {Leefer}}, \bibinfo {author} {\bibfnamefont {C.~T.~M.}\ \bibnamefont
  {Weber}}, \bibinfo {author} {\bibfnamefont {A.}~\bibnamefont {Cing{\"o}z}},
  \bibinfo {author} {\bibfnamefont {J.~R.}\ \bibnamefont {Torgerson}},\ and\
  \bibinfo {author} {\bibfnamefont {D.}~\bibnamefont {Budker}},\ }\bibfield
  {title} {\bibinfo {title} {{New Limits on Variation of the Fine-Structure
  Constant Using Atomic Dysprosium}},\ }\href
  {https://doi.org/10.1103/PhysRevLett.111.060801} {\bibfield  {journal}
  {\bibinfo  {journal} {Phys. Rev. Lett.}\ }\textbf {\bibinfo {volume} {111}},\
  \bibinfo {pages} {060801} (\bibinfo {year} {2013})},\ \Eprint
  {https://arxiv.org/abs/1304.6940} {arXiv:1304.6940} \BibitemShut {NoStop}%
\bibitem [{\citenamefont {Godun}\ \emph {et~al.}(2014)\citenamefont {Godun},
  \citenamefont {Nisbet-Jones}, \citenamefont {Jones}, \citenamefont {King},
  \citenamefont {Johnson}, \citenamefont {Margolis}, \citenamefont {Szymaniec},
  \citenamefont {Lea}, \citenamefont {Bongs},\ and\ \citenamefont
  {Gill}}]{Godun:2014naa}%
  \BibitemOpen
  \bibfield  {author} {\bibinfo {author} {\bibfnamefont {R.~M.}\ \bibnamefont
  {Godun}}, \bibinfo {author} {\bibfnamefont {P.~B.~R.}\ \bibnamefont
  {Nisbet-Jones}}, \bibinfo {author} {\bibfnamefont {J.~M.}\ \bibnamefont
  {Jones}}, \bibinfo {author} {\bibfnamefont {S.~A.}\ \bibnamefont {King}},
  \bibinfo {author} {\bibfnamefont {L.~A.~M.}\ \bibnamefont {Johnson}},
  \bibinfo {author} {\bibfnamefont {H.~S.}\ \bibnamefont {Margolis}}, \bibinfo
  {author} {\bibfnamefont {K.}~\bibnamefont {Szymaniec}}, \bibinfo {author}
  {\bibfnamefont {S.~N.}\ \bibnamefont {Lea}}, \bibinfo {author} {\bibfnamefont
  {K.}~\bibnamefont {Bongs}},\ and\ \bibinfo {author} {\bibfnamefont
  {P.}~\bibnamefont {Gill}},\ }\bibfield  {title} {\bibinfo {title} {{Frequency
  Ratio of Two Optical Clock Transitions in {${}^{171}$Yb$^{+}$} and
  Constraints on the Time Variation of Fundamental Constants}},\ }\href
  {https://doi.org/10.1103/PhysRevLett.113.210801} {\bibfield  {journal}
  {\bibinfo  {journal} {Phys. Rev. Lett.}\ }\textbf {\bibinfo {volume} {113}},\
  \bibinfo {pages} {210801} (\bibinfo {year} {2014})},\ \Eprint
  {https://arxiv.org/abs/1407.0164} {arXiv:1407.0164} \BibitemShut {NoStop}%
\bibitem [{\citenamefont {Huntemann}\ \emph {et~al.}(2014)\citenamefont
  {Huntemann}, \citenamefont {Lipphardt}, \citenamefont {Tamm}, \citenamefont
  {Gerginov}, \citenamefont {Weyers},\ and\ \citenamefont
  {Peik}}]{Huntemann:2014dya}%
  \BibitemOpen
  \bibfield  {author} {\bibinfo {author} {\bibfnamefont {N.}~\bibnamefont
  {Huntemann}}, \bibinfo {author} {\bibfnamefont {B.}~\bibnamefont
  {Lipphardt}}, \bibinfo {author} {\bibfnamefont {C.}~\bibnamefont {Tamm}},
  \bibinfo {author} {\bibfnamefont {V.}~\bibnamefont {Gerginov}}, \bibinfo
  {author} {\bibfnamefont {S.}~\bibnamefont {Weyers}},\ and\ \bibinfo {author}
  {\bibfnamefont {E.}~\bibnamefont {Peik}},\ }\bibfield  {title} {\bibinfo
  {title} {{Improved Limit on a Temporal Variation of $m_{p}/m_{e}$ from
  Comparisons of {Yb$^{+}$} and {Cs} Atomic Clocks}},\ }\href
  {https://doi.org/10.1103/PhysRevLett.113.210802} {\bibfield  {journal}
  {\bibinfo  {journal} {Phys. Rev. Lett.}\ }\textbf {\bibinfo {volume} {113}},\
  \bibinfo {pages} {210802} (\bibinfo {year} {2014})},\ \Eprint
  {https://arxiv.org/abs/1407.4408} {arXiv:1407.4408} \BibitemShut {NoStop}%
\bibitem [{\citenamefont {McGrew}\ \emph {et~al.}(2019)\citenamefont {McGrew},
  \citenamefont {Zhang}, \citenamefont {Leopardi}, \citenamefont {Fasano},
  \citenamefont {Nicolodi}, \citenamefont {Beloy}, \citenamefont {Yao},
  \citenamefont {Sherman}, \citenamefont {Sch{\"a}ffer}, \citenamefont
  {Savory}, \citenamefont {Brown}, \citenamefont {R{\"o}misch}, \citenamefont
  {Oates}, \citenamefont {Parker}, \citenamefont {Fortier},\ and\ \citenamefont
  {Ludlow}}]{McGrew2018}%
  \BibitemOpen
  \bibfield  {author} {\bibinfo {author} {\bibfnamefont {W.~F.}\ \bibnamefont
  {McGrew}}, \bibinfo {author} {\bibfnamefont {X.}~\bibnamefont {Zhang}},
  \bibinfo {author} {\bibfnamefont {H.}~\bibnamefont {Leopardi}}, \bibinfo
  {author} {\bibfnamefont {R.~J.}\ \bibnamefont {Fasano}}, \bibinfo {author}
  {\bibfnamefont {D.}~\bibnamefont {Nicolodi}}, \bibinfo {author}
  {\bibfnamefont {K.}~\bibnamefont {Beloy}}, \bibinfo {author} {\bibfnamefont
  {J.}~\bibnamefont {Yao}}, \bibinfo {author} {\bibfnamefont {J.~A.}\
  \bibnamefont {Sherman}}, \bibinfo {author} {\bibfnamefont {S.~A.}\
  \bibnamefont {Sch{\"a}ffer}}, \bibinfo {author} {\bibfnamefont
  {J.}~\bibnamefont {Savory}}, \bibinfo {author} {\bibfnamefont {R.~C.}\
  \bibnamefont {Brown}}, \bibinfo {author} {\bibfnamefont {S.}~\bibnamefont
  {R{\"o}misch}}, \bibinfo {author} {\bibfnamefont {C.~W.}\ \bibnamefont
  {Oates}}, \bibinfo {author} {\bibfnamefont {T.~E.}\ \bibnamefont {Parker}},
  \bibinfo {author} {\bibfnamefont {T.~M.}\ \bibnamefont {Fortier}},\ and\
  \bibinfo {author} {\bibfnamefont {A.~D.}\ \bibnamefont {Ludlow}},\ }\bibfield
   {title} {\bibinfo {title} {Towards the optical second:\ verifying optical
  clocks at the {{SI}} limit},\ }\href
  {https://doi.org/10.1364/OPTICA.6.000448} {\bibfield  {journal} {\bibinfo
  {journal} {Optica}\ }\textbf {\bibinfo {volume} {6}},\ \bibinfo {pages} {448}
  (\bibinfo {year} {2019})},\ \Eprint {https://arxiv.org/abs/1811.05885}
  {arXiv:1811.05885} \BibitemShut {NoStop}%
\bibitem [{\citenamefont {Lange}\ \emph {et~al.}(2021)\citenamefont {Lange},
  \citenamefont {Huntemann}, \citenamefont {Rahm}, \citenamefont {Sanner},
  \citenamefont {Shao}, \citenamefont {Lipphardt}, \citenamefont {Tamm},
  \citenamefont {Weyers},\ and\ \citenamefont {Peik}}]{Lange:2020cul}%
  \BibitemOpen
  \bibfield  {author} {\bibinfo {author} {\bibfnamefont {R.}~\bibnamefont
  {Lange}}, \bibinfo {author} {\bibfnamefont {N.}~\bibnamefont {Huntemann}},
  \bibinfo {author} {\bibfnamefont {J.~M.}\ \bibnamefont {Rahm}}, \bibinfo
  {author} {\bibfnamefont {C.}~\bibnamefont {Sanner}}, \bibinfo {author}
  {\bibfnamefont {H.}~\bibnamefont {Shao}}, \bibinfo {author} {\bibfnamefont
  {B.}~\bibnamefont {Lipphardt}}, \bibinfo {author} {\bibfnamefont {{\relax
  Chr}.}~\bibnamefont {Tamm}}, \bibinfo {author} {\bibfnamefont
  {S.}~\bibnamefont {Weyers}},\ and\ \bibinfo {author} {\bibfnamefont
  {E.}~\bibnamefont {Peik}},\ }\bibfield  {title} {\bibinfo {title} {Improved
  {{Limits}} for {{Violations}} of {{Local Position Invariance}} from {{Atomic
  Clock Comparisons}}},\ }\href
  {https://doi.org/10.1103/PhysRevLett.126.011102} {\bibfield  {journal}
  {\bibinfo  {journal} {Phys. Rev. Lett.}\ }\textbf {\bibinfo {volume} {126}},\
  \bibinfo {pages} {011102} (\bibinfo {year} {2021})},\ \Eprint
  {https://arxiv.org/abs/2010.06620} {arXiv:2010.06620} \BibitemShut {NoStop}%
\bibitem [{\citenamefont {Schwarz}\ \emph {et~al.}(2020)\citenamefont
  {Schwarz}, \citenamefont {D{\"o}rscher}, \citenamefont {{Al-Masoudi}},
  \citenamefont {Benkler}, \citenamefont {Legero}, \citenamefont {Sterr},
  \citenamefont {Weyers}, \citenamefont {Rahm}, \citenamefont {Lipphardt},\
  and\ \citenamefont {Lisdat}}]{Schwarz:2020upg}%
  \BibitemOpen
  \bibfield  {author} {\bibinfo {author} {\bibfnamefont {R.}~\bibnamefont
  {Schwarz}}, \bibinfo {author} {\bibfnamefont {S.}~\bibnamefont
  {D{\"o}rscher}}, \bibinfo {author} {\bibfnamefont {A.}~\bibnamefont
  {{Al-Masoudi}}}, \bibinfo {author} {\bibfnamefont {E.}~\bibnamefont
  {Benkler}}, \bibinfo {author} {\bibfnamefont {T.}~\bibnamefont {Legero}},
  \bibinfo {author} {\bibfnamefont {U.}~\bibnamefont {Sterr}}, \bibinfo
  {author} {\bibfnamefont {S.}~\bibnamefont {Weyers}}, \bibinfo {author}
  {\bibfnamefont {J.}~\bibnamefont {Rahm}}, \bibinfo {author} {\bibfnamefont
  {B.}~\bibnamefont {Lipphardt}},\ and\ \bibinfo {author} {\bibfnamefont
  {C.}~\bibnamefont {Lisdat}},\ }\bibfield  {title} {\bibinfo {title} {Long
  term measurement of the 87 {{Sr}} clock frequency at the limit of primary
  {{Cs}} clocks},\ }\href {https://doi.org/10.1103/PhysRevResearch.2.033242}
  {\bibfield  {journal} {\bibinfo  {journal} {Phys. Rev. Res.}\ }\textbf
  {\bibinfo {volume} {2}},\ \bibinfo {pages} {033242} (\bibinfo {year}
  {2020})},\ \Eprint {https://arxiv.org/abs/2005.07408} {arXiv:2005.07408}
  \BibitemShut {NoStop}%
\bibitem [{\citenamefont {Filzinger}\ \emph {et~al.}(2023)\citenamefont
  {Filzinger}, \citenamefont {D{\"o}rscher}, \citenamefont {Lange},
  \citenamefont {Klose}, \citenamefont {Steinel}, \citenamefont {Benkler},
  \citenamefont {Peik}, \citenamefont {Lisdat},\ and\ \citenamefont
  {Huntemann}}]{Filzinger:2023zrs}%
  \BibitemOpen
  \bibfield  {author} {\bibinfo {author} {\bibfnamefont {M.}~\bibnamefont
  {Filzinger}}, \bibinfo {author} {\bibfnamefont {S.}~\bibnamefont
  {D{\"o}rscher}}, \bibinfo {author} {\bibfnamefont {R.}~\bibnamefont {Lange}},
  \bibinfo {author} {\bibfnamefont {J.}~\bibnamefont {Klose}}, \bibinfo
  {author} {\bibfnamefont {M.}~\bibnamefont {Steinel}}, \bibinfo {author}
  {\bibfnamefont {E.}~\bibnamefont {Benkler}}, \bibinfo {author} {\bibfnamefont
  {E.}~\bibnamefont {Peik}}, \bibinfo {author} {\bibfnamefont {C.}~\bibnamefont
  {Lisdat}},\ and\ \bibinfo {author} {\bibfnamefont {N.}~\bibnamefont
  {Huntemann}},\ }\bibfield  {title} {\bibinfo {title} {{Improved Limits on the
  Coupling of Ultralight Bosonic Dark Matter to Photons from Optical Atomic
  Clock Comparisons}},\ }\href {https://doi.org/10.1103/PhysRevLett.130.253001}
  {\bibfield  {journal} {\bibinfo  {journal} {Phys. Rev. Lett.}\ }\textbf
  {\bibinfo {volume} {130}},\ \bibinfo {pages} {253001} (\bibinfo {year}
  {2023})},\ \Eprint {https://arxiv.org/abs/2301.03433} {arXiv:2301.03433}
  \BibitemShut {NoStop}%
\bibitem [{\citenamefont {Derevianko}\ and\ \citenamefont
  {Pospelov}(2014)}]{Derevianko:2013oaa}%
  \BibitemOpen
  \bibfield  {author} {\bibinfo {author} {\bibfnamefont {A.}~\bibnamefont
  {Derevianko}}\ and\ \bibinfo {author} {\bibfnamefont {M.}~\bibnamefont
  {Pospelov}},\ }\bibfield  {title} {\bibinfo {title} {Hunting for topological
  dark matter with atomic clocks},\ }\href {https://doi.org/10.1038/nphys3137}
  {\bibfield  {journal} {\bibinfo  {journal} {Nat. Phys.}\ }\textbf {\bibinfo
  {volume} {10}},\ \bibinfo {pages} {933} (\bibinfo {year} {2014})},\ \Eprint
  {https://arxiv.org/abs/1311.1244} {arXiv:1311.1244} \BibitemShut {NoStop}%
\bibitem [{\citenamefont {Wcis{\l}o}\ \emph {et~al.}(2016)\citenamefont
  {Wcis{\l}o}, \citenamefont {Morzy{\'n}ski}, \citenamefont {Bober},
  \citenamefont {Cygan}, \citenamefont {Lisak}, \citenamefont {Ciury{\l}o},\
  and\ \citenamefont {Zawada}}]{Wcislo:2016qng}%
  \BibitemOpen
  \bibfield  {author} {\bibinfo {author} {\bibfnamefont {P.}~\bibnamefont
  {Wcis{\l}o}}, \bibinfo {author} {\bibfnamefont {P.}~\bibnamefont
  {Morzy{\'n}ski}}, \bibinfo {author} {\bibfnamefont {M.}~\bibnamefont
  {Bober}}, \bibinfo {author} {\bibfnamefont {A.}~\bibnamefont {Cygan}},
  \bibinfo {author} {\bibfnamefont {D.}~\bibnamefont {Lisak}}, \bibinfo
  {author} {\bibfnamefont {R.}~\bibnamefont {Ciury{\l}o}},\ and\ \bibinfo
  {author} {\bibfnamefont {M.}~\bibnamefont {Zawada}},\ }\bibfield  {title}
  {\bibinfo {title} {Experimental constraint on dark matter detection with
  optical atomic clocks},\ }\href {https://doi.org/10.1038/s41550-016-0009}
  {\bibfield  {journal} {\bibinfo  {journal} {Nat. Astron.}\ }\textbf {\bibinfo
  {volume} {1}},\ \bibinfo {pages} {0009} (\bibinfo {year} {2016})},\ \Eprint
  {https://arxiv.org/abs/1605.05763} {arXiv:1605.05763} \BibitemShut {NoStop}%
\bibitem [{\citenamefont {Roberts}\ \emph {et~al.}(2017)\citenamefont
  {Roberts}, \citenamefont {Blewitt}, \citenamefont {Dailey}, \citenamefont
  {Murphy}, \citenamefont {Pospelov}, \citenamefont {Rollings}, \citenamefont
  {Sherman}, \citenamefont {Williams},\ and\ \citenamefont
  {Derevianko}}]{Roberts:2017hla}%
  \BibitemOpen
  \bibfield  {author} {\bibinfo {author} {\bibfnamefont {B.~M.}\ \bibnamefont
  {Roberts}}, \bibinfo {author} {\bibfnamefont {G.}~\bibnamefont {Blewitt}},
  \bibinfo {author} {\bibfnamefont {C.}~\bibnamefont {Dailey}}, \bibinfo
  {author} {\bibfnamefont {M.}~\bibnamefont {Murphy}}, \bibinfo {author}
  {\bibfnamefont {M.}~\bibnamefont {Pospelov}}, \bibinfo {author}
  {\bibfnamefont {A.}~\bibnamefont {Rollings}}, \bibinfo {author}
  {\bibfnamefont {J.}~\bibnamefont {Sherman}}, \bibinfo {author} {\bibfnamefont
  {W.}~\bibnamefont {Williams}},\ and\ \bibinfo {author} {\bibfnamefont
  {A.}~\bibnamefont {Derevianko}},\ }\bibfield  {title} {\bibinfo {title}
  {Search for domain wall dark matter with atomic clocks on board global
  positioning system satellites},\ }\href
  {https://doi.org/10.1038/s41467-017-01440-4} {\bibfield  {journal} {\bibinfo
  {journal} {Nat. Commun.}\ }\textbf {\bibinfo {volume} {8}},\ \bibinfo {pages}
  {1195} (\bibinfo {year} {2017})},\ \Eprint {https://arxiv.org/abs/1704.06844}
  {arXiv:1704.06844} \BibitemShut {NoStop}%
\bibitem [{\citenamefont {Roberts}\ and\ \citenamefont
  {Derevianko}(2021)}]{Roberts:2018agv}%
  \BibitemOpen
  \bibfield  {author} {\bibinfo {author} {\bibfnamefont {B.~M.}\ \bibnamefont
  {Roberts}}\ and\ \bibinfo {author} {\bibfnamefont {A.}~\bibnamefont
  {Derevianko}},\ }\bibfield  {title} {\bibinfo {title} {Precision
  {{Measurement Noise Asymmetry}} and {{Its Annual Modulation}} as a {{Dark
  Matter Signature}}},\ }\href {https://doi.org/10.3390/universe7030050}
  {\bibfield  {journal} {\bibinfo  {journal} {Universe}\ }\textbf {\bibinfo
  {volume} {7}},\ \bibinfo {pages} {50} (\bibinfo {year} {2021})},\ \Eprint
  {https://arxiv.org/abs/1803.00617} {arXiv:1803.00617} \BibitemShut {NoStop}%
\bibitem [{\citenamefont {Wcis{\l}o}\ \emph {et~al.}(2018)\citenamefont
  {Wcis{\l}o} \emph {et~al.}}]{Wcislo:2018ojh}%
  \BibitemOpen
  \bibfield  {author} {\bibinfo {author} {\bibfnamefont {P.}~\bibnamefont
  {Wcis{\l}o}} \emph {et~al.},\ }\bibfield  {title} {\bibinfo {title} {New
  bounds on dark matter coupling from a global network of optical atomic
  clocks},\ }\href {https://doi.org/10.1126/sciadv.aau4869} {\bibfield
  {journal} {\bibinfo  {journal} {Sci. Adv.}\ }\textbf {\bibinfo {volume}
  {4}},\ \bibinfo {pages} {eaau4869} (\bibinfo {year} {2018})},\ \Eprint
  {https://arxiv.org/abs/1806.04762} {arXiv:1806.04762} \BibitemShut {NoStop}%
\bibitem [{\citenamefont {Roberts}\ \emph {et~al.}(2020)\citenamefont {Roberts}
  \emph {et~al.}}]{Roberts:2019sfo}%
  \BibitemOpen
  \bibfield  {author} {\bibinfo {author} {\bibfnamefont {B.~M.}\ \bibnamefont
  {Roberts}} \emph {et~al.},\ }\bibfield  {title} {\bibinfo {title} {Search for
  transient variations of the fine structure constant and dark matter using
  fiber-linked optical atomic clocks},\ }\href
  {https://doi.org/10.1088/1367-2630/abaace} {\bibfield  {journal} {\bibinfo
  {journal} {N. J. Phys.}\ }\textbf {\bibinfo {volume} {22}},\ \bibinfo {pages}
  {093010} (\bibinfo {year} {2020})},\ \Eprint
  {https://arxiv.org/abs/1907.02661} {arXiv:1907.02661} \BibitemShut {NoStop}%
\bibitem [{\citenamefont {Van~Tilburg}\ \emph {et~al.}(2015)\citenamefont
  {Van~Tilburg}, \citenamefont {Leefer}, \citenamefont {Bougas},\ and\
  \citenamefont {Budker}}]{VanTilburg:2015oza}%
  \BibitemOpen
  \bibfield  {author} {\bibinfo {author} {\bibfnamefont {K.}~\bibnamefont
  {Van~Tilburg}}, \bibinfo {author} {\bibfnamefont {N.}~\bibnamefont {Leefer}},
  \bibinfo {author} {\bibfnamefont {L.}~\bibnamefont {Bougas}},\ and\ \bibinfo
  {author} {\bibfnamefont {D.}~\bibnamefont {Budker}},\ }\bibfield  {title}
  {\bibinfo {title} {Search for {{Ultralight Scalar Dark Matter}} with {{Atomic
  Spectroscopy}}},\ }\href {https://doi.org/10.1103/PhysRevLett.115.011802}
  {\bibfield  {journal} {\bibinfo  {journal} {Phys. Rev. Lett.}\ }\textbf
  {\bibinfo {volume} {115}},\ \bibinfo {pages} {011802} (\bibinfo {year}
  {2015})},\ \Eprint {https://arxiv.org/abs/1503.06886} {arXiv:1503.06886}
  \BibitemShut {NoStop}%
\bibitem [{\citenamefont {Hees}\ \emph {et~al.}(2016)\citenamefont {Hees},
  \citenamefont {Gu{\'e}na}, \citenamefont {Abgrall}, \citenamefont {Bize},\
  and\ \citenamefont {Wolf}}]{Hees:2016gop}%
  \BibitemOpen
  \bibfield  {author} {\bibinfo {author} {\bibfnamefont {A.}~\bibnamefont
  {Hees}}, \bibinfo {author} {\bibfnamefont {J.}~\bibnamefont {Gu{\'e}na}},
  \bibinfo {author} {\bibfnamefont {M.}~\bibnamefont {Abgrall}}, \bibinfo
  {author} {\bibfnamefont {S.}~\bibnamefont {Bize}},\ and\ \bibinfo {author}
  {\bibfnamefont {P.}~\bibnamefont {Wolf}},\ }\bibfield  {title} {\bibinfo
  {title} {Searching for an {{Oscillating Massive Scalar Field}} as a {{Dark
  Matter Candidate Using Atomic Hyperfine Frequency Comparisons}}},\ }\href
  {https://doi.org/10.1103/PhysRevLett.117.061301} {\bibfield  {journal}
  {\bibinfo  {journal} {Phys. Rev. Lett.}\ }\textbf {\bibinfo {volume} {117}},\
  \bibinfo {pages} {061301} (\bibinfo {year} {2016})},\ \Eprint
  {https://arxiv.org/abs/1604.08514} {arXiv:1604.08514} \BibitemShut {NoStop}%
\bibitem [{\citenamefont {Antypas}\ \emph {et~al.}(2019)\citenamefont
  {Antypas}, \citenamefont {Tretiak}, \citenamefont {Garcon}, \citenamefont
  {Ozeri}, \citenamefont {Perez},\ and\ \citenamefont
  {Budker}}]{Antypas:2019qji}%
  \BibitemOpen
  \bibfield  {author} {\bibinfo {author} {\bibfnamefont {D.}~\bibnamefont
  {Antypas}}, \bibinfo {author} {\bibfnamefont {O.}~\bibnamefont {Tretiak}},
  \bibinfo {author} {\bibfnamefont {A.}~\bibnamefont {Garcon}}, \bibinfo
  {author} {\bibfnamefont {R.}~\bibnamefont {Ozeri}}, \bibinfo {author}
  {\bibfnamefont {G.}~\bibnamefont {Perez}},\ and\ \bibinfo {author}
  {\bibfnamefont {D.}~\bibnamefont {Budker}},\ }\bibfield  {title} {\bibinfo
  {title} {Scalar {{Dark Matter}} in the {{Radio-Frequency Band}}:\
  {{Atomic-Spectroscopy Search Results}}},\ }\href
  {https://doi.org/10.1103/PhysRevLett.123.141102} {\bibfield  {journal}
  {\bibinfo  {journal} {Phys. Rev. Lett.}\ }\textbf {\bibinfo {volume} {123}},\
  \bibinfo {pages} {141102} (\bibinfo {year} {2019})},\ \Eprint
  {https://arxiv.org/abs/1905.02968} {arXiv:1905.02968} \BibitemShut {NoStop}%
\bibitem [{\citenamefont {Savalle}\ \emph {et~al.}(2021)\citenamefont
  {Savalle}, \citenamefont {Hees}, \citenamefont {Frank}, \citenamefont
  {Cantin}, \citenamefont {Pottie}, \citenamefont {Roberts}, \citenamefont
  {Cros}, \citenamefont {McAllister},\ and\ \citenamefont
  {Wolf}}]{Savalle:2020vgz}%
  \BibitemOpen
  \bibfield  {author} {\bibinfo {author} {\bibfnamefont {E.}~\bibnamefont
  {Savalle}}, \bibinfo {author} {\bibfnamefont {A.}~\bibnamefont {Hees}},
  \bibinfo {author} {\bibfnamefont {F.}~\bibnamefont {Frank}}, \bibinfo
  {author} {\bibfnamefont {E.}~\bibnamefont {Cantin}}, \bibinfo {author}
  {\bibfnamefont {P.-E.}\ \bibnamefont {Pottie}}, \bibinfo {author}
  {\bibfnamefont {B.~M.}\ \bibnamefont {Roberts}}, \bibinfo {author}
  {\bibfnamefont {L.}~\bibnamefont {Cros}}, \bibinfo {author} {\bibfnamefont
  {B.~T.}\ \bibnamefont {McAllister}},\ and\ \bibinfo {author} {\bibfnamefont
  {P.}~\bibnamefont {Wolf}},\ }\bibfield  {title} {\bibinfo {title} {Searching
  for {{Dark Matter}} with an {{Optical Cavity}} and an {{Unequal-Delay
  Interferometer}}},\ }\href {https://doi.org/10.1103/PhysRevLett.126.051301}
  {\bibfield  {journal} {\bibinfo  {journal} {Phys. Rev. Lett.}\ }\textbf
  {\bibinfo {volume} {126}},\ \bibinfo {pages} {051301} (\bibinfo {year}
  {2021})},\ \Eprint {https://arxiv.org/abs/2006.07055} {arXiv:2006.07055}
  \BibitemShut {NoStop}%
\bibitem [{\citenamefont {Kennedy}\ \emph {et~al.}(2020)\citenamefont
  {Kennedy}, \citenamefont {Oelker}, \citenamefont {Robinson}, \citenamefont
  {Bothwell}, \citenamefont {Kedar}, \citenamefont {Milner}, \citenamefont
  {Marti}, \citenamefont {Derevianko},\ and\ \citenamefont
  {Ye}}]{Kennedy:2020bac}%
  \BibitemOpen
  \bibfield  {author} {\bibinfo {author} {\bibfnamefont {C.~J.}\ \bibnamefont
  {Kennedy}}, \bibinfo {author} {\bibfnamefont {E.}~\bibnamefont {Oelker}},
  \bibinfo {author} {\bibfnamefont {J.~M.}\ \bibnamefont {Robinson}}, \bibinfo
  {author} {\bibfnamefont {T.}~\bibnamefont {Bothwell}}, \bibinfo {author}
  {\bibfnamefont {D.}~\bibnamefont {Kedar}}, \bibinfo {author} {\bibfnamefont
  {W.~R.}\ \bibnamefont {Milner}}, \bibinfo {author} {\bibfnamefont {G.~E.}\
  \bibnamefont {Marti}}, \bibinfo {author} {\bibfnamefont {A.}~\bibnamefont
  {Derevianko}},\ and\ \bibinfo {author} {\bibfnamefont {J.}~\bibnamefont
  {Ye}},\ }\bibfield  {title} {\bibinfo {title} {Precision {{Metrology Meets
  Cosmology}}:\ {{Improved Constraints}} on {{Ultralight Dark Matter}} from
  {{Atom-Cavity Frequency Comparisons}}},\ }\href
  {https://doi.org/10.1103/PhysRevLett.125.201302} {\bibfield  {journal}
  {\bibinfo  {journal} {Phys. Rev. Lett.}\ }\textbf {\bibinfo {volume} {125}},\
  \bibinfo {pages} {201302} (\bibinfo {year} {2020})},\ \Eprint
  {https://arxiv.org/abs/2008.08773} {arXiv:2008.08773} \BibitemShut {NoStop}%
\bibitem [{\citenamefont {{K.~Beloy {\sl et al.} (The BACON
  Collaboration)}}(2021)}]{Beloy:2020tgz}%
  \BibitemOpen
  \bibfield  {author} {\bibinfo {author} {\bibnamefont {{K.~Beloy {\sl et al.}
  (The BACON Collaboration)}}},\ }\bibfield  {title} {\bibinfo {title}
  {Frequency ratio measurements at 18-digit accuracy using an optical clock
  network},\ }\href {https://doi.org/10.1038/s41586-021-03253-4} {\bibfield
  {journal} {\bibinfo  {journal} {Nature}\ }\textbf {\bibinfo {volume} {591}},\
  \bibinfo {pages} {564} (\bibinfo {year} {2021})},\ \Eprint
  {https://arxiv.org/abs/2005.14694} {arXiv:2005.14694} \BibitemShut {NoStop}%
\bibitem [{\citenamefont {Oswald}\ \emph {et~al.}(2022)\citenamefont {Oswald},
  \citenamefont {Nevsky}, \citenamefont {Vogt}, \citenamefont {Schiller},
  \citenamefont {Figueroa}, \citenamefont {Zhang}, \citenamefont {Tretiak},
  \citenamefont {Antypas}, \citenamefont {Budker}, \citenamefont {Banerjee},\
  and\ \citenamefont {Perez}}]{Oswald:2021vtc}%
  \BibitemOpen
  \bibfield  {author} {\bibinfo {author} {\bibfnamefont {R.}~\bibnamefont
  {Oswald}}, \bibinfo {author} {\bibfnamefont {A.}~\bibnamefont {Nevsky}},
  \bibinfo {author} {\bibfnamefont {V.}~\bibnamefont {Vogt}}, \bibinfo {author}
  {\bibfnamefont {S.}~\bibnamefont {Schiller}}, \bibinfo {author}
  {\bibfnamefont {N.~L.}\ \bibnamefont {Figueroa}}, \bibinfo {author}
  {\bibfnamefont {K.}~\bibnamefont {Zhang}}, \bibinfo {author} {\bibfnamefont
  {O.}~\bibnamefont {Tretiak}}, \bibinfo {author} {\bibfnamefont
  {D.}~\bibnamefont {Antypas}}, \bibinfo {author} {\bibfnamefont
  {D.}~\bibnamefont {Budker}}, \bibinfo {author} {\bibfnamefont
  {A.}~\bibnamefont {Banerjee}},\ and\ \bibinfo {author} {\bibfnamefont
  {G.}~\bibnamefont {Perez}},\ }\bibfield  {title} {\bibinfo {title} {Search
  for {{Dark-Matter-Induced Oscillations}} of {{Fundamental Constants Using
  Molecular Spectroscopy}}},\ }\href
  {https://doi.org/10.1103/PhysRevLett.129.031302} {\bibfield  {journal}
  {\bibinfo  {journal} {Phys. Rev. Lett.}\ }\textbf {\bibinfo {volume} {129}},\
  \bibinfo {pages} {031302} (\bibinfo {year} {2022})},\ \Eprint
  {https://arxiv.org/abs/2111.06883} {arXiv:2111.06883} \BibitemShut {NoStop}%
\bibitem [{\citenamefont {Tretiak}\ \emph {et~al.}(2022)\citenamefont
  {Tretiak}, \citenamefont {Zhang}, \citenamefont {Figueroa}, \citenamefont
  {Antypas}, \citenamefont {Brogna}, \citenamefont {Banerjee}, \citenamefont
  {Perez},\ and\ \citenamefont {Budker}}]{Tretiak:2022ndx}%
  \BibitemOpen
  \bibfield  {author} {\bibinfo {author} {\bibfnamefont {O.}~\bibnamefont
  {Tretiak}}, \bibinfo {author} {\bibfnamefont {X.}~\bibnamefont {Zhang}},
  \bibinfo {author} {\bibfnamefont {N.~L.}\ \bibnamefont {Figueroa}}, \bibinfo
  {author} {\bibfnamefont {D.}~\bibnamefont {Antypas}}, \bibinfo {author}
  {\bibfnamefont {A.}~\bibnamefont {Brogna}}, \bibinfo {author} {\bibfnamefont
  {A.}~\bibnamefont {Banerjee}}, \bibinfo {author} {\bibfnamefont
  {G.}~\bibnamefont {Perez}},\ and\ \bibinfo {author} {\bibfnamefont
  {D.}~\bibnamefont {Budker}},\ }\bibfield  {title} {\bibinfo {title} {Improved
  {{Bounds}} on {{Ultralight Scalar Dark Matter}} in the {{Radio-Frequency
  Range}}},\ }\href {https://doi.org/10.1103/PhysRevLett.129.031301} {\bibfield
   {journal} {\bibinfo  {journal} {Phys. Rev. Lett.}\ }\textbf {\bibinfo
  {volume} {129}},\ \bibinfo {pages} {031301} (\bibinfo {year} {2022})},\
  \Eprint {https://arxiv.org/abs/2201.02042} {arXiv:2201.02042} \BibitemShut
  {NoStop}%
\bibitem [{\citenamefont {Zhang}\ \emph {et~al.}(2023)\citenamefont {Zhang},
  \citenamefont {Banerjee}, \citenamefont {Leyser}, \citenamefont {Perez},
  \citenamefont {Schiller}, \citenamefont {Budker},\ and\ \citenamefont
  {Antypas}}]{Zhang:2022ewz}%
  \BibitemOpen
  \bibfield  {author} {\bibinfo {author} {\bibfnamefont {X.}~\bibnamefont
  {Zhang}}, \bibinfo {author} {\bibfnamefont {A.}~\bibnamefont {Banerjee}},
  \bibinfo {author} {\bibfnamefont {M.}~\bibnamefont {Leyser}}, \bibinfo
  {author} {\bibfnamefont {G.}~\bibnamefont {Perez}}, \bibinfo {author}
  {\bibfnamefont {S.}~\bibnamefont {Schiller}}, \bibinfo {author}
  {\bibfnamefont {D.}~\bibnamefont {Budker}},\ and\ \bibinfo {author}
  {\bibfnamefont {D.}~\bibnamefont {Antypas}},\ }\bibfield  {title} {\bibinfo
  {title} {Search for {{Ultralight Dark Matter}} with {{Spectroscopy}} of
  {{Radio-Frequency Atomic Transitions}}},\ }\href
  {https://doi.org/10.1103/PhysRevLett.130.251002} {\bibfield  {journal}
  {\bibinfo  {journal} {Phys. Rev. Lett.}\ }\textbf {\bibinfo {volume} {130}},\
  \bibinfo {pages} {251002} (\bibinfo {year} {2023})},\ \Eprint
  {https://arxiv.org/abs/2212.04413} {arXiv:2212.04413} \BibitemShut {NoStop}%
\bibitem [{\citenamefont {Kobayashi}\ \emph {et~al.}(2022)\citenamefont
  {Kobayashi}, \citenamefont {Takamizawa}, \citenamefont {Akamatsu},
  \citenamefont {Kawasaki}, \citenamefont {Nishiyama}, \citenamefont {Hosaka},
  \citenamefont {Hisai}, \citenamefont {Wada}, \citenamefont {Inaba},
  \citenamefont {Tanabe},\ and\ \citenamefont {Yasuda}}]{Kobayashi:2022vsf}%
  \BibitemOpen
  \bibfield  {author} {\bibinfo {author} {\bibfnamefont {T.}~\bibnamefont
  {Kobayashi}}, \bibinfo {author} {\bibfnamefont {A.}~\bibnamefont
  {Takamizawa}}, \bibinfo {author} {\bibfnamefont {D.}~\bibnamefont
  {Akamatsu}}, \bibinfo {author} {\bibfnamefont {A.}~\bibnamefont {Kawasaki}},
  \bibinfo {author} {\bibfnamefont {A.}~\bibnamefont {Nishiyama}}, \bibinfo
  {author} {\bibfnamefont {K.}~\bibnamefont {Hosaka}}, \bibinfo {author}
  {\bibfnamefont {Y.}~\bibnamefont {Hisai}}, \bibinfo {author} {\bibfnamefont
  {M.}~\bibnamefont {Wada}}, \bibinfo {author} {\bibfnamefont {H.}~\bibnamefont
  {Inaba}}, \bibinfo {author} {\bibfnamefont {T.}~\bibnamefont {Tanabe}},\ and\
  \bibinfo {author} {\bibfnamefont {M.}~\bibnamefont {Yasuda}},\ }\bibfield
  {title} {\bibinfo {title} {Search for {{Ultralight Dark Matter}} from
  {{Long-Term Frequency Comparisons}} of {{Optical}} and {{Microwave Atomic
  Clocks}}},\ }\href {https://doi.org/10.1103/PhysRevLett.129.241301}
  {\bibfield  {journal} {\bibinfo  {journal} {Phys. Rev. Lett.}\ }\textbf
  {\bibinfo {volume} {129}},\ \bibinfo {pages} {241301} (\bibinfo {year}
  {2022})},\ \Eprint {https://arxiv.org/abs/2212.05721} {arXiv:2212.05721}
  \BibitemShut {NoStop}%
\bibitem [{\citenamefont {Sherrill}\ \emph {et~al.}(2023)\citenamefont
  {Sherrill}, \citenamefont {Parsons}, \citenamefont {Baynham}, \citenamefont
  {Bowden}, \citenamefont {Anne~Curtis}, \citenamefont {Hendricks},
  \citenamefont {Hill}, \citenamefont {Hobson}, \citenamefont {Margolis},
  \citenamefont {Robertson}, \citenamefont {Schioppo}, \citenamefont
  {Szymaniec}, \citenamefont {Tofful}, \citenamefont {Tunesi}, \citenamefont
  {Godun},\ and\ \citenamefont {Calmet}}]{Sherrill:2023zah}%
  \BibitemOpen
  \bibfield  {author} {\bibinfo {author} {\bibfnamefont {N.}~\bibnamefont
  {Sherrill}}, \bibinfo {author} {\bibfnamefont {A.~O.}\ \bibnamefont
  {Parsons}}, \bibinfo {author} {\bibfnamefont {C.~F.~A.}\ \bibnamefont
  {Baynham}}, \bibinfo {author} {\bibfnamefont {W.}~\bibnamefont {Bowden}},
  \bibinfo {author} {\bibfnamefont {E.}~\bibnamefont {Anne~Curtis}}, \bibinfo
  {author} {\bibfnamefont {R.}~\bibnamefont {Hendricks}}, \bibinfo {author}
  {\bibfnamefont {I.~R.}\ \bibnamefont {Hill}}, \bibinfo {author}
  {\bibfnamefont {R.}~\bibnamefont {Hobson}}, \bibinfo {author} {\bibfnamefont
  {H.~S.}\ \bibnamefont {Margolis}}, \bibinfo {author} {\bibfnamefont {B.~I.}\
  \bibnamefont {Robertson}}, \bibinfo {author} {\bibfnamefont {M.}~\bibnamefont
  {Schioppo}}, \bibinfo {author} {\bibfnamefont {K.}~\bibnamefont {Szymaniec}},
  \bibinfo {author} {\bibfnamefont {A.}~\bibnamefont {Tofful}}, \bibinfo
  {author} {\bibfnamefont {J.}~\bibnamefont {Tunesi}}, \bibinfo {author}
  {\bibfnamefont {R.~M.}\ \bibnamefont {Godun}},\ and\ \bibinfo {author}
  {\bibfnamefont {X.}~\bibnamefont {Calmet}},\ }\bibfield  {title} {\bibinfo
  {title} {Analysis of atomic-clock data to constrain variations of fundamental
  constants},\ }\href {https://doi.org/10.1088/1367-2630/aceff6} {\bibfield
  {journal} {\bibinfo  {journal} {N. J. Phys.}\ }\textbf {\bibinfo {volume}
  {25}},\ \bibinfo {pages} {093012} (\bibinfo {year} {2023})},\ \Eprint
  {https://arxiv.org/abs/2302.04565} {arXiv:2302.04565} \BibitemShut {NoStop}%
\bibitem [{\citenamefont {Stadnik}\ and\ \citenamefont
  {Flambaum}(2015{\natexlab{b}})}]{Stadnik:2014tta}%
  \BibitemOpen
  \bibfield  {author} {\bibinfo {author} {\bibfnamefont {Y.~V.}\ \bibnamefont
  {Stadnik}}\ and\ \bibinfo {author} {\bibfnamefont {V.~V.}\ \bibnamefont
  {Flambaum}},\ }\bibfield  {title} {\bibinfo {title} {{Searching for Dark
  Matter and Variation of Fundamental Constants with Laser and Maser
  Interferometry}},\ }\href {https://doi.org/10.1103/PhysRevLett.114.161301}
  {\bibfield  {journal} {\bibinfo  {journal} {Phys. Rev. Lett.}\ }\textbf
  {\bibinfo {volume} {114}},\ \bibinfo {pages} {161301} (\bibinfo {year}
  {2015}{\natexlab{b}})},\ \Eprint {https://arxiv.org/abs/1412.7801}
  {arXiv:1412.7801} \BibitemShut {NoStop}%
\bibitem [{\citenamefont {Geraci}\ \emph {et~al.}(2019)\citenamefont {Geraci},
  \citenamefont {Bradley}, \citenamefont {Gao}, \citenamefont {Weinstein},\
  and\ \citenamefont {Derevianko}}]{Geraci:2018fax}%
  \BibitemOpen
  \bibfield  {author} {\bibinfo {author} {\bibfnamefont {A.~A.}\ \bibnamefont
  {Geraci}}, \bibinfo {author} {\bibfnamefont {C.}~\bibnamefont {Bradley}},
  \bibinfo {author} {\bibfnamefont {D.}~\bibnamefont {Gao}}, \bibinfo {author}
  {\bibfnamefont {J.}~\bibnamefont {Weinstein}},\ and\ \bibinfo {author}
  {\bibfnamefont {A.}~\bibnamefont {Derevianko}},\ }\bibfield  {title}
  {\bibinfo {title} {Searching for {{Ultralight Dark Matter}} with {{Optical
  Cavities}}},\ }\href {https://doi.org/10.1103/PhysRevLett.123.031304}
  {\bibfield  {journal} {\bibinfo  {journal} {Phys. Rev. Lett.}\ }\textbf
  {\bibinfo {volume} {123}},\ \bibinfo {pages} {031304} (\bibinfo {year}
  {2019})},\ \Eprint {https://arxiv.org/abs/1808.00540} {arXiv:1808.00540}
  \BibitemShut {NoStop}%
\bibitem [{\citenamefont {Schioppo}\ \emph {et~al.}(2022)\citenamefont
  {Schioppo}, \citenamefont {Kronjäger}, \citenamefont {Silva}, \citenamefont
  {Ilieva}, \citenamefont {Paterson}, \citenamefont {Baynham}, \citenamefont
  {Bowden}, \citenamefont {Hill}, \citenamefont {Hobson}, \citenamefont
  {Vianello}, \citenamefont {Dovale-Álvarez}, \citenamefont {Williams},
  \citenamefont {Marra}, \citenamefont {Margolis}, \citenamefont {Amy-Klein},
  \citenamefont {Lopez}, \citenamefont {Cantin}, \citenamefont {Álvarez
  Martínez}, \citenamefont {Le~Targat}, \citenamefont {Pottie}, \citenamefont
  {Quintin}, \citenamefont {Legero}, \citenamefont {Häfner}, \citenamefont
  {Sterr}, \citenamefont {Schwarz}, \citenamefont {Dörscher}, \citenamefont
  {Lisdat}, \citenamefont {Koke}, \citenamefont {Kuhl}, \citenamefont
  {Waterholter}, \citenamefont {Benkler},\ and\ \citenamefont
  {Grosche}}]{Schioppo:2022iqe}%
  \BibitemOpen
  \bibfield  {author} {\bibinfo {author} {\bibfnamefont {M.}~\bibnamefont
  {Schioppo}}, \bibinfo {author} {\bibfnamefont {J.}~\bibnamefont
  {Kronjäger}}, \bibinfo {author} {\bibfnamefont {A.}~\bibnamefont {Silva}},
  \bibinfo {author} {\bibfnamefont {R.}~\bibnamefont {Ilieva}}, \bibinfo
  {author} {\bibfnamefont {J.~W.}\ \bibnamefont {Paterson}}, \bibinfo {author}
  {\bibfnamefont {C.~F.~A.}\ \bibnamefont {Baynham}}, \bibinfo {author}
  {\bibfnamefont {W.}~\bibnamefont {Bowden}}, \bibinfo {author} {\bibfnamefont
  {I.~R.}\ \bibnamefont {Hill}}, \bibinfo {author} {\bibfnamefont
  {R.}~\bibnamefont {Hobson}}, \bibinfo {author} {\bibfnamefont
  {A.}~\bibnamefont {Vianello}}, \bibinfo {author} {\bibfnamefont
  {M.}~\bibnamefont {Dovale-Álvarez}}, \bibinfo {author} {\bibfnamefont
  {R.~A.}\ \bibnamefont {Williams}}, \bibinfo {author} {\bibfnamefont
  {G.}~\bibnamefont {Marra}}, \bibinfo {author} {\bibfnamefont {H.~S.}\
  \bibnamefont {Margolis}}, \bibinfo {author} {\bibfnamefont {A.}~\bibnamefont
  {Amy-Klein}}, \bibinfo {author} {\bibfnamefont {O.}~\bibnamefont {Lopez}},
  \bibinfo {author} {\bibfnamefont {E.}~\bibnamefont {Cantin}}, \bibinfo
  {author} {\bibfnamefont {H.}~\bibnamefont {Álvarez Martínez}}, \bibinfo
  {author} {\bibfnamefont {R.}~\bibnamefont {Le~Targat}}, \bibinfo {author}
  {\bibfnamefont {P.~E.}\ \bibnamefont {Pottie}}, \bibinfo {author}
  {\bibfnamefont {N.}~\bibnamefont {Quintin}}, \bibinfo {author} {\bibfnamefont
  {T.}~\bibnamefont {Legero}}, \bibinfo {author} {\bibfnamefont
  {S.}~\bibnamefont {Häfner}}, \bibinfo {author} {\bibfnamefont
  {U.}~\bibnamefont {Sterr}}, \bibinfo {author} {\bibfnamefont
  {R.}~\bibnamefont {Schwarz}}, \bibinfo {author} {\bibfnamefont
  {S.}~\bibnamefont {Dörscher}}, \bibinfo {author} {\bibfnamefont
  {C.}~\bibnamefont {Lisdat}}, \bibinfo {author} {\bibfnamefont
  {S.}~\bibnamefont {Koke}}, \bibinfo {author} {\bibfnamefont {A.}~\bibnamefont
  {Kuhl}}, \bibinfo {author} {\bibfnamefont {T.}~\bibnamefont {Waterholter}},
  \bibinfo {author} {\bibfnamefont {E.}~\bibnamefont {Benkler}},\ and\ \bibinfo
  {author} {\bibfnamefont {G.}~\bibnamefont {Grosche}},\ }\bibfield  {title}
  {\bibinfo {title} {Comparing ultrastable lasers at $7 \times 10^{-17}$
  fractional frequency instability through a 2220 km optical fibre network},\
  }\href {https://doi.org/10.1038/s41467-021-27884-3} {\bibfield  {journal}
  {\bibinfo  {journal} {Nat. Commun.}\ }\textbf {\bibinfo {volume} {13}},\
  \bibinfo {pages} {212} (\bibinfo {year} {2022})}\BibitemShut {NoStop}%
\bibitem [{\citenamefont {Hui}\ \emph {et~al.}(2017)\citenamefont {Hui},
  \citenamefont {Ostriker}, \citenamefont {Tremaine},\ and\ \citenamefont
  {Witten}}]{Hui:2016ltb}%
  \BibitemOpen
  \bibfield  {author} {\bibinfo {author} {\bibfnamefont {L.}~\bibnamefont
  {Hui}}, \bibinfo {author} {\bibfnamefont {J.~P.}\ \bibnamefont {Ostriker}},
  \bibinfo {author} {\bibfnamefont {S.}~\bibnamefont {Tremaine}},\ and\
  \bibinfo {author} {\bibfnamefont {E.}~\bibnamefont {Witten}},\ }\bibfield
  {title} {\bibinfo {title} {Ultralight scalars as cosmological dark matter},\
  }\href {https://doi.org/10.1103/PhysRevD.95.043541} {\bibfield  {journal}
  {\bibinfo  {journal} {Phys. Rev. D}\ }\textbf {\bibinfo {volume} {95}},\
  \bibinfo {pages} {043541} (\bibinfo {year} {2017})},\ \Eprint
  {https://arxiv.org/abs/1610.08297} {arXiv:1610.08297} \BibitemShut {NoStop}%
\bibitem [{\citenamefont {Hees}\ \emph {et~al.}(2018)\citenamefont {Hees},
  \citenamefont {Minazzoli}, \citenamefont {Savalle}, \citenamefont {Stadnik},\
  and\ \citenamefont {Wolf}}]{Hees:2018fpg}%
  \BibitemOpen
  \bibfield  {author} {\bibinfo {author} {\bibfnamefont {A.}~\bibnamefont
  {Hees}}, \bibinfo {author} {\bibfnamefont {O.}~\bibnamefont {Minazzoli}},
  \bibinfo {author} {\bibfnamefont {E.}~\bibnamefont {Savalle}}, \bibinfo
  {author} {\bibfnamefont {Y.~V.}\ \bibnamefont {Stadnik}},\ and\ \bibinfo
  {author} {\bibfnamefont {P.}~\bibnamefont {Wolf}},\ }\bibfield  {title}
  {\bibinfo {title} {Violation of the equivalence principle from light scalar
  dark matter},\ }\href {https://doi.org/10.1103/PhysRevD.98.064051} {\bibfield
   {journal} {\bibinfo  {journal} {Phys. Rev. D}\ }\textbf {\bibinfo {volume}
  {98}},\ \bibinfo {pages} {064051} (\bibinfo {year} {2018})},\ \Eprint
  {https://arxiv.org/abs/1807.04512} {arXiv:1807.04512} \BibitemShut {NoStop}%
\bibitem [{\citenamefont {Freese}\ \emph {et~al.}(2013)\citenamefont {Freese},
  \citenamefont {Lisanti},\ and\ \citenamefont {Savage}}]{Freese:2012xd}%
  \BibitemOpen
  \bibfield  {author} {\bibinfo {author} {\bibfnamefont {K.}~\bibnamefont
  {Freese}}, \bibinfo {author} {\bibfnamefont {M.}~\bibnamefont {Lisanti}},\
  and\ \bibinfo {author} {\bibfnamefont {C.}~\bibnamefont {Savage}},\
  }\bibfield  {title} {\bibinfo {title} {{Colloquium:\ {{Annual}} Modulation of
  Dark Matter}},\ }\href {https://doi.org/10.1103/RevModPhys.85.1561}
  {\bibfield  {journal} {\bibinfo  {journal} {Rev. Mod. Phys.}\ }\textbf
  {\bibinfo {volume} {85}},\ \bibinfo {pages} {1561} (\bibinfo {year}
  {2013})},\ \Eprint {https://arxiv.org/abs/1209.3339} {arxiv:1209.3339}
  \BibitemShut {NoStop}%
\bibitem [{\citenamefont {Graham}\ and\ \citenamefont
  {Rajendran}(2011)}]{Graham:2011qk}%
  \BibitemOpen
  \bibfield  {author} {\bibinfo {author} {\bibfnamefont {P.~W.}\ \bibnamefont
  {Graham}}\ and\ \bibinfo {author} {\bibfnamefont {S.}~\bibnamefont
  {Rajendran}},\ }\bibfield  {title} {\bibinfo {title} {Axion dark matter
  detection with cold molecules},\ }\href
  {https://doi.org/10.1103/PhysRevD.84.055013} {\bibfield  {journal} {\bibinfo
  {journal} {Phys. Rev. D}\ }\textbf {\bibinfo {volume} {84}},\ \bibinfo
  {pages} {055013} (\bibinfo {year} {2011})},\ \Eprint
  {https://arxiv.org/abs/1101.2691} {arXiv:1101.2691} \BibitemShut {NoStop}%
\bibitem [{\citenamefont {Damour}\ and\ \citenamefont
  {Donoghue}(2010)}]{Damour:2010rp}%
  \BibitemOpen
  \bibfield  {author} {\bibinfo {author} {\bibfnamefont {T.}~\bibnamefont
  {Damour}}\ and\ \bibinfo {author} {\bibfnamefont {J.~F.}\ \bibnamefont
  {Donoghue}},\ }\bibfield  {title} {\bibinfo {title} {Equivalence principle
  violations and couplings of a light dilaton},\ }\href
  {https://doi.org/10.1103/PhysRevD.82.084033} {\bibfield  {journal} {\bibinfo
  {journal} {Phys. Rev. D}\ }\textbf {\bibinfo {volume} {82}},\ \bibinfo
  {pages} {084033} (\bibinfo {year} {2010})},\ \Eprint
  {https://arxiv.org/abs/1007.2792} {arXiv:1007.2792} \BibitemShut {NoStop}%
\bibitem [{\citenamefont {Kim}\ \emph {et~al.}(2024)\citenamefont {Kim},
  \citenamefont {Lenoci}, \citenamefont {Perez},\ and\ \citenamefont
  {Ratzinger}}]{Kim:2023pvt}%
  \BibitemOpen
  \bibfield  {author} {\bibinfo {author} {\bibfnamefont {H.}~\bibnamefont
  {Kim}}, \bibinfo {author} {\bibfnamefont {A.}~\bibnamefont {Lenoci}},
  \bibinfo {author} {\bibfnamefont {G.}~\bibnamefont {Perez}},\ and\ \bibinfo
  {author} {\bibfnamefont {W.}~\bibnamefont {Ratzinger}},\ }\bibfield  {title}
  {\bibinfo {title} {Probing an ultralight {{QCD}} axion with electromagnetic
  quadratic interaction},\ }\href {https://doi.org/10.1103/PhysRevD.109.015030}
  {\bibfield  {journal} {\bibinfo  {journal} {Phys. Rev. D}\ }\textbf {\bibinfo
  {volume} {109}},\ \bibinfo {pages} {015030} (\bibinfo {year} {2024})},\
  \Eprint {https://arxiv.org/abs/2307.14962} {arXiv:2307.14962} \BibitemShut
  {NoStop}%
\bibitem [{\citenamefont {Banerjee}\ \emph {et~al.}(2023)\citenamefont
  {Banerjee}, \citenamefont {Perez}, \citenamefont {Safronova}, \citenamefont
  {Savoray},\ and\ \citenamefont {Shalit}}]{Banerjee:2022sqg}%
  \BibitemOpen
  \bibfield  {author} {\bibinfo {author} {\bibfnamefont {A.}~\bibnamefont
  {Banerjee}}, \bibinfo {author} {\bibfnamefont {G.}~\bibnamefont {Perez}},
  \bibinfo {author} {\bibfnamefont {M.}~\bibnamefont {Safronova}}, \bibinfo
  {author} {\bibfnamefont {I.}~\bibnamefont {Savoray}},\ and\ \bibinfo {author}
  {\bibfnamefont {A.}~\bibnamefont {Shalit}},\ }\bibfield  {title} {\bibinfo
  {title} {The phenomenology of quadratically coupled ultra light dark
  matter},\ }\href {https://doi.org/10.1007/JHEP10(2023)042} {\bibfield
  {journal} {\bibinfo  {journal} {J. High Energy Phys.}\ }\textbf {\bibinfo
  {volume} {10}}\bibfield  {number} {\bibinfo  {number} { (2023)},\ \bibinfo
  {pages} {42}},\ }\Eprint {https://arxiv.org/abs/2211.05174}
  {arXiv:2211.05174} \BibitemShut {NoStop}%
\bibitem [{\citenamefont {Flambaum}\ and\ \citenamefont
  {Samsonov}(2023)}]{Flambaum:2023bnw}%
  \BibitemOpen
  \bibfield  {author} {\bibinfo {author} {\bibfnamefont {V.~V.}\ \bibnamefont
  {Flambaum}}\ and\ \bibinfo {author} {\bibfnamefont {I.~B.}\ \bibnamefont
  {Samsonov}},\ }\bibfield  {title} {\bibinfo {title} {Fluctuations of atomic
  energy levels due to axion and scalar fields},\ }\href
  {https://doi.org/10.1103/PhysRevD.108.075022} {\bibfield  {journal} {\bibinfo
   {journal} {Phys. Rev. D}\ }\textbf {\bibinfo {volume} {108}},\ \bibinfo
  {pages} {075022} (\bibinfo {year} {2023})},\ \Eprint
  {https://arxiv.org/abs/2302.11167} {arXiv:2302.11167} \BibitemShut {NoStop}%
\bibitem [{\citenamefont {Beadle}\ \emph {et~al.}()\citenamefont {Beadle},
  \citenamefont {Ellis}, \citenamefont {Quevillon},\ and\ \citenamefont
  {Vuong}}]{Beadle:2023flm}%
  \BibitemOpen
  \bibfield  {author} {\bibinfo {author} {\bibfnamefont {C.}~\bibnamefont
  {Beadle}}, \bibinfo {author} {\bibfnamefont {S.~A.~R.}\ \bibnamefont
  {Ellis}}, \bibinfo {author} {\bibfnamefont {J.}~\bibnamefont {Quevillon}},\
  and\ \bibinfo {author} {\bibfnamefont {P.~N.~H.}\ \bibnamefont {Vuong}},\
  }\bibfield  {title} {\bibinfo {title} {Quadratic {{Coupling}} of the
  {{Axion}} to {{Photons}}},\ }\href {http://arxiv.org/abs/2307.10362} {\
  }\Eprint {https://arxiv.org/abs/2307.10362} {arXiv:2307.10362} \BibitemShut
  {NoStop}%
\bibitem [{\citenamefont {Savalle}\ \emph {et~al.}()\citenamefont {Savalle},
  \citenamefont {Roberts}, \citenamefont {Frank}, \citenamefont {Pottie},
  \citenamefont {McAllister}, \citenamefont {Dailey}, \citenamefont
  {Derevianko},\ and\ \citenamefont {Wolf}}]{Savalle:2019jsb}%
  \BibitemOpen
  \bibfield  {author} {\bibinfo {author} {\bibfnamefont {E.}~\bibnamefont
  {Savalle}}, \bibinfo {author} {\bibfnamefont {B.~M.}\ \bibnamefont
  {Roberts}}, \bibinfo {author} {\bibfnamefont {F.}~\bibnamefont {Frank}},
  \bibinfo {author} {\bibfnamefont {P.-E.}\ \bibnamefont {Pottie}}, \bibinfo
  {author} {\bibfnamefont {B.~T.}\ \bibnamefont {McAllister}}, \bibinfo
  {author} {\bibfnamefont {C.}~\bibnamefont {Dailey}}, \bibinfo {author}
  {\bibfnamefont {A.}~\bibnamefont {Derevianko}},\ and\ \bibinfo {author}
  {\bibfnamefont {P.}~\bibnamefont {Wolf}},\ }\bibfield  {title} {\bibinfo
  {title} {Novel approaches to dark-matter detection using space-time separated
  clocks},\ }\href {https://arxiv.org/abs/1902.07192} {\ }\Eprint
  {https://arxiv.org/abs/1902.07192} {arXiv:1902.07192} \BibitemShut {NoStop}%
\bibitem [{\citenamefont {Dzuba}\ \emph {et~al.}(1999)\citenamefont {Dzuba},
  \citenamefont {Flambaum},\ and\ \citenamefont {Webb}}]{Dzuba:1999zz}%
  \BibitemOpen
  \bibfield  {author} {\bibinfo {author} {\bibfnamefont {V.~A.}\ \bibnamefont
  {Dzuba}}, \bibinfo {author} {\bibfnamefont {V.~V.}\ \bibnamefont
  {Flambaum}},\ and\ \bibinfo {author} {\bibfnamefont {J.~K.}\ \bibnamefont
  {Webb}},\ }\bibfield  {title} {\bibinfo {title} {{Space-Time Variation of
  Physical Constants and Relativistic Corrections in Atoms}},\ }\href
  {https://doi.org/10.1103/PhysRevLett.82.888} {\bibfield  {journal} {\bibinfo
  {journal} {Phys. Rev. Lett.}\ }\textbf {\bibinfo {volume} {82}},\ \bibinfo
  {pages} {888} (\bibinfo {year} {1999})},\ \Eprint
  {https://arxiv.org/abs/physics/9802029} {arXiv:physics/9802029} \BibitemShut
  {NoStop}%
\bibitem [{\citenamefont {Flambaum}\ \emph {et~al.}(2004)\citenamefont
  {Flambaum}, \citenamefont {Leinweber}, \citenamefont {Thomas},\ and\
  \citenamefont {Young}}]{Flambaum:2004tm}%
  \BibitemOpen
  \bibfield  {author} {\bibinfo {author} {\bibfnamefont {V.~V.}\ \bibnamefont
  {Flambaum}}, \bibinfo {author} {\bibfnamefont {D.~B.}\ \bibnamefont
  {Leinweber}}, \bibinfo {author} {\bibfnamefont {A.~W.}\ \bibnamefont
  {Thomas}},\ and\ \bibinfo {author} {\bibfnamefont {R.~D.}\ \bibnamefont
  {Young}},\ }\bibfield  {title} {\bibinfo {title} {Limits on variations of the
  quark masses, {{QCD}} scale, and fine structure constant},\ }\href
  {https://doi.org/10.1103/PhysRevD.69.115006} {\bibfield  {journal} {\bibinfo
  {journal} {Phys. Rev. D}\ }\textbf {\bibinfo {volume} {69}},\ \bibinfo
  {pages} {115006} (\bibinfo {year} {2004})}\BibitemShut {NoStop}%
\bibitem [{\citenamefont {Flambaum}\ and\ \citenamefont
  {Tedesco}(2006)}]{Flambaum:2006ip}%
  \BibitemOpen
  \bibfield  {author} {\bibinfo {author} {\bibfnamefont {V.~V.}\ \bibnamefont
  {Flambaum}}\ and\ \bibinfo {author} {\bibfnamefont {A.~F.}\ \bibnamefont
  {Tedesco}},\ }\bibfield  {title} {\bibinfo {title} {Dependence of nuclear
  magnetic moments on quark masses and limits on temporal variation of
  fundamental constants from atomic clock experiments},\ }\href
  {https://doi.org/10.1103/PhysRevC.73.055501} {\bibfield  {journal} {\bibinfo
  {journal} {Phys. Rev. C}\ }\textbf {\bibinfo {volume} {73}},\ \bibinfo
  {pages} {055501} (\bibinfo {year} {2006})},\ \Eprint
  {https://arxiv.org/abs/nucl-th/0601050} {arXiv:nucl-th/0601050} \BibitemShut
  {NoStop}%
\bibitem [{\citenamefont {Banerjee}\ \emph {et~al.}()\citenamefont {Banerjee},
  \citenamefont {Budker}, \citenamefont {Filzinger}, \citenamefont {Huntemann},
  \citenamefont {Paz}, \citenamefont {Perez}, \citenamefont {Porsev},\ and\
  \citenamefont {Safronova}}]{Banerjee:2023bjc}%
  \BibitemOpen
  \bibfield  {author} {\bibinfo {author} {\bibfnamefont {A.}~\bibnamefont
  {Banerjee}}, \bibinfo {author} {\bibfnamefont {D.}~\bibnamefont {Budker}},
  \bibinfo {author} {\bibfnamefont {M.}~\bibnamefont {Filzinger}}, \bibinfo
  {author} {\bibfnamefont {N.}~\bibnamefont {Huntemann}}, \bibinfo {author}
  {\bibfnamefont {G.}~\bibnamefont {Paz}}, \bibinfo {author} {\bibfnamefont
  {G.}~\bibnamefont {Perez}}, \bibinfo {author} {\bibfnamefont
  {S.}~\bibnamefont {Porsev}},\ and\ \bibinfo {author} {\bibfnamefont
  {M.}~\bibnamefont {Safronova}},\ }\bibfield  {title} {\bibinfo {title}
  {Oscillating nuclear charge radii as sensors for ultralight dark matter},\
  }\href {http://arxiv.org/abs/2301.10784} {\ }\Eprint
  {https://arxiv.org/abs/2301.10784} {arXiv:2301.10784} \BibitemShut {NoStop}%
\bibitem [{\citenamefont {Flambaum}\ and\ \citenamefont
  {Mansour}(2023)}]{Flambaum:2023drb}%
  \BibitemOpen
  \bibfield  {author} {\bibinfo {author} {\bibfnamefont {V.~V.}\ \bibnamefont
  {Flambaum}}\ and\ \bibinfo {author} {\bibfnamefont {A.~J.}\ \bibnamefont
  {Mansour}},\ }\bibfield  {title} {\bibinfo {title} {Variation of the
  {{Quadrupole Hyperfine Structure}} and {{Nuclear Radius}} due to an
  {{Interaction}} with {{Scalar}} and {{Axion Dark Matter}}},\ }\href
  {https://doi.org/10.1103/PhysRevLett.131.113004} {\bibfield  {journal}
  {\bibinfo  {journal} {Phys. Rev. Lett.}\ }\textbf {\bibinfo {volume} {131}},\
  \bibinfo {pages} {113004} (\bibinfo {year} {2023})},\ \Eprint
  {https://arxiv.org/abs/2304.04469} {arXiv:2304.04469} \BibitemShut {NoStop}%
\bibitem [{\citenamefont {Kozlov}\ and\ \citenamefont
  {Budker}(2019)}]{Kozlov:2018qid}%
  \BibitemOpen
  \bibfield  {author} {\bibinfo {author} {\bibfnamefont {M.~G.}\ \bibnamefont
  {Kozlov}}\ and\ \bibinfo {author} {\bibfnamefont {D.}~\bibnamefont
  {Budker}},\ }\bibfield  {title} {\bibinfo {title} {Comment on {{Sensitivity
  Coefficients}} to {{Variation}} of {{Fundamental Constants}}},\ }\href
  {https://doi.org/10.1002/andp.201800254} {\bibfield  {journal} {\bibinfo
  {journal} {Ann. Phys.}\ }\textbf {\bibinfo {volume} {531}},\ \bibinfo {pages}
  {1800254} (\bibinfo {year} {2019})},\ \Eprint
  {https://arxiv.org/abs/1807.08337} {arXiv:1807.08337} \BibitemShut {NoStop}%
\bibitem [{\citenamefont {Antypas}\ \emph {et~al.}(2020)\citenamefont
  {Antypas}, \citenamefont {Budker}, \citenamefont {Flambaum}, \citenamefont
  {Kozlov}, \citenamefont {Perez},\ and\ \citenamefont {Ye}}]{Antypas:2019yvv}%
  \BibitemOpen
  \bibfield  {author} {\bibinfo {author} {\bibfnamefont {D.}~\bibnamefont
  {Antypas}}, \bibinfo {author} {\bibfnamefont {D.}~\bibnamefont {Budker}},
  \bibinfo {author} {\bibfnamefont {V.~V.}\ \bibnamefont {Flambaum}}, \bibinfo
  {author} {\bibfnamefont {M.~G.}\ \bibnamefont {Kozlov}}, \bibinfo {author}
  {\bibfnamefont {G.}~\bibnamefont {Perez}},\ and\ \bibinfo {author}
  {\bibfnamefont {J.}~\bibnamefont {Ye}},\ }\bibfield  {title} {\bibinfo
  {title} {Fast {{Apparent Oscillations}} of {{Fundamental Constants}}},\
  }\href {https://doi.org/10.1002/andp.201900566} {\bibfield  {journal}
  {\bibinfo  {journal} {Ann. Phys.}\ }\textbf {\bibinfo {volume} {532}},\
  \bibinfo {pages} {1900566} (\bibinfo {year} {2020})},\ \Eprint
  {https://arxiv.org/abs/1912.01335} {arXiv:1912.01335} \BibitemShut {NoStop}%
\bibitem [{\citenamefont {Lopez}\ \emph {et~al.}(2012)\citenamefont {Lopez},
  \citenamefont {Haboucha}, \citenamefont {Chanteau}, \citenamefont
  {Chardonnet}, \citenamefont {{Amy-Klein}},\ and\ \citenamefont
  {Santarelli}}]{Lopez:2012}%
  \BibitemOpen
  \bibfield  {author} {\bibinfo {author} {\bibfnamefont {O.}~\bibnamefont
  {Lopez}}, \bibinfo {author} {\bibfnamefont {A.}~\bibnamefont {Haboucha}},
  \bibinfo {author} {\bibfnamefont {B.}~\bibnamefont {Chanteau}}, \bibinfo
  {author} {\bibfnamefont {C.}~\bibnamefont {Chardonnet}}, \bibinfo {author}
  {\bibfnamefont {A.}~\bibnamefont {{Amy-Klein}}},\ and\ \bibinfo {author}
  {\bibfnamefont {G.}~\bibnamefont {Santarelli}},\ }\bibfield  {title}
  {\bibinfo {title} {Ultra-stable long distance optical frequency distribution
  using the {{Internet}} fiber network},\ }\href
  {https://doi.org/10.1364/OE.20.023518} {\bibfield  {journal} {\bibinfo
  {journal} {Opt. Express}\ }\textbf {\bibinfo {volume} {20}},\ \bibinfo
  {pages} {23518} (\bibinfo {year} {2012})}\BibitemShut {NoStop}%
\bibitem [{\citenamefont {Raupach}\ \emph {et~al.}(2015)\citenamefont
  {Raupach}, \citenamefont {Koczwara},\ and\ \citenamefont
  {Grosche}}]{Raupach:2015}%
  \BibitemOpen
  \bibfield  {author} {\bibinfo {author} {\bibfnamefont {S.~M.~F.}\
  \bibnamefont {Raupach}}, \bibinfo {author} {\bibfnamefont {A.}~\bibnamefont
  {Koczwara}},\ and\ \bibinfo {author} {\bibfnamefont {G.}~\bibnamefont
  {Grosche}},\ }\bibfield  {title} {\bibinfo {title} {Brillouin amplification
  supports $1\times 10^{-20}$ uncertainty in optical frequency transfer over
  1400 km of underground fiber},\ }\href
  {https://doi.org/10.1103/PhysRevA.92.021801} {\bibfield  {journal} {\bibinfo
  {journal} {Phys. Rev. A}\ }\textbf {\bibinfo {volume} {92}},\ \bibinfo
  {pages} {021801(R)} (\bibinfo {year} {2015})}\BibitemShut {NoStop}%
\bibitem [{\citenamefont {Chiodo}\ \emph {et~al.}(2015)\citenamefont {Chiodo},
  \citenamefont {Quintin}, \citenamefont {Stefani}, \citenamefont {Wiotte},
  \citenamefont {Camisard}, \citenamefont {Chardonnet}, \citenamefont
  {Santarelli}, \citenamefont {{Amy-Klein}}, \citenamefont {Pottie},\ and\
  \citenamefont {Lopez}}]{Chiodo:2015oma}%
  \BibitemOpen
  \bibfield  {author} {\bibinfo {author} {\bibfnamefont {N.}~\bibnamefont
  {Chiodo}}, \bibinfo {author} {\bibfnamefont {N.}~\bibnamefont {Quintin}},
  \bibinfo {author} {\bibfnamefont {F.}~\bibnamefont {Stefani}}, \bibinfo
  {author} {\bibfnamefont {F.}~\bibnamefont {Wiotte}}, \bibinfo {author}
  {\bibfnamefont {E.}~\bibnamefont {Camisard}}, \bibinfo {author}
  {\bibfnamefont {C.}~\bibnamefont {Chardonnet}}, \bibinfo {author}
  {\bibfnamefont {G.}~\bibnamefont {Santarelli}}, \bibinfo {author}
  {\bibfnamefont {A.}~\bibnamefont {{Amy-Klein}}}, \bibinfo {author}
  {\bibfnamefont {P.-E.}\ \bibnamefont {Pottie}},\ and\ \bibinfo {author}
  {\bibfnamefont {O.}~\bibnamefont {Lopez}},\ }\bibfield  {title} {\bibinfo
  {title} {Cascaded optical fiber link using the internet network for remote
  clocks comparison},\ }\href {https://doi.org/10.1364/OE.23.033927} {\bibfield
   {journal} {\bibinfo  {journal} {Opt. Express}\ }\textbf {\bibinfo {volume}
  {23}},\ \bibinfo {pages} {33927} (\bibinfo {year} {2015})}\BibitemShut
  {NoStop}%
\bibitem [{\citenamefont {Lisdat}\ \emph {et~al.}(2016)\citenamefont {Lisdat}
  \emph {et~al.}}]{Lisdat2016}%
  \BibitemOpen
  \bibfield  {author} {\bibinfo {author} {\bibfnamefont {C.}~\bibnamefont
  {Lisdat}} \emph {et~al.},\ }\bibfield  {title} {\bibinfo {title} {A clock
  network for geodesy and fundamental science},\ }\href
  {https://doi.org/10.1038/ncomms12443} {\bibfield  {journal} {\bibinfo
  {journal} {Nat. Commun.}\ }\textbf {\bibinfo {volume} {7}},\ \bibinfo {pages}
  {12443} (\bibinfo {year} {2016})},\ \Eprint
  {https://arxiv.org/abs/1511.07735} {arXiv:1511.07735} \BibitemShut {NoStop}%
\bibitem [{\citenamefont {Koke}\ \emph {et~al.}(2019)\citenamefont {Koke},
  \citenamefont {Kuhl}, \citenamefont {Waterholter}, \citenamefont {Raupach},
  \citenamefont {Lopez}, \citenamefont {Cantin}, \citenamefont {Quintin},
  \citenamefont {{Amy-Klein}}, \citenamefont {Pottie},\ and\ \citenamefont
  {Grosche}}]{Koke:2019bwo}%
  \BibitemOpen
  \bibfield  {author} {\bibinfo {author} {\bibfnamefont {S.}~\bibnamefont
  {Koke}}, \bibinfo {author} {\bibfnamefont {A.}~\bibnamefont {Kuhl}}, \bibinfo
  {author} {\bibfnamefont {T.}~\bibnamefont {Waterholter}}, \bibinfo {author}
  {\bibfnamefont {S.~M.~F.}\ \bibnamefont {Raupach}}, \bibinfo {author}
  {\bibfnamefont {O.}~\bibnamefont {Lopez}}, \bibinfo {author} {\bibfnamefont
  {E.}~\bibnamefont {Cantin}}, \bibinfo {author} {\bibfnamefont
  {N.}~\bibnamefont {Quintin}}, \bibinfo {author} {\bibfnamefont
  {A.}~\bibnamefont {{Amy-Klein}}}, \bibinfo {author} {\bibfnamefont {P.-E.}\
  \bibnamefont {Pottie}},\ and\ \bibinfo {author} {\bibfnamefont
  {G.}~\bibnamefont {Grosche}},\ }\bibfield  {title} {\bibinfo {title}
  {Combining fiber {{Brillouin}} amplification with a repeater laser station
  for fiber-based optical frequency dissemination over 1400 km},\ }\href
  {https://doi.org/10.1088/1367-2630/ab5d95} {\bibfield  {journal} {\bibinfo
  {journal} {N. J. Phys.}\ }\textbf {\bibinfo {volume} {21}},\ \bibinfo {pages}
  {123017} (\bibinfo {year} {2019})},\ \Eprint
  {https://arxiv.org/abs/1911.01215} {arXiv:1911.01215} \BibitemShut {NoStop}%
\bibitem [{\citenamefont {Clivati}\ \emph {et~al.}(2022)\citenamefont {Clivati}
  \emph {et~al.}}]{ITALY2022}%
  \BibitemOpen
  \bibfield  {author} {\bibinfo {author} {\bibfnamefont {C.}~\bibnamefont
  {Clivati}} \emph {et~al.},\ }\bibfield  {title} {\bibinfo {title} {{Coherent
  Optical-Fiber Link Across Italy and France}},\ }\href
  {https://doi.org/10.1103/PhysRevApplied.18.054009} {\bibfield  {journal}
  {\bibinfo  {journal} {Phys. Rev. Appl.}\ }\textbf {\bibinfo {volume} {18}},\
  \bibinfo {pages} {054009} (\bibinfo {year} {2022})}\BibitemShut {NoStop}%
\bibitem [{\citenamefont {Delva}\ \emph
  {et~al.}(2017{\natexlab{a}})\citenamefont {Delva} \emph
  {et~al.}}]{Delva:2017lie}%
  \BibitemOpen
  \bibfield  {author} {\bibinfo {author} {\bibfnamefont {P.}~\bibnamefont
  {Delva}} \emph {et~al.},\ }\bibfield  {title} {\bibinfo {title} {Test of
  {{Special Relativity Using}} a {{Fiber Network}} of {{Optical Clocks}}},\
  }\href {https://doi.org/10.1103/PhysRevLett.118.221102} {\bibfield  {journal}
  {\bibinfo  {journal} {Phys. Rev. Lett.}\ }\textbf {\bibinfo {volume} {118}},\
  \bibinfo {pages} {221102} (\bibinfo {year} {2017}{\natexlab{a}})},\ \Eprint
  {https://arxiv.org/abs/1703.04426} {arXiv:1703.04426} \BibitemShut {NoStop}%
\bibitem [{\citenamefont {Schioppo}\ \emph {et~al.}(2021)\citenamefont
  {Schioppo} \emph {et~al.}}]{SchioppoZenodo:2021}%
  \BibitemOpen
  \bibfield  {author} {\bibinfo {author} {\bibfnamefont {M.}~\bibnamefont
  {Schioppo}} \emph {et~al.},\ }\href {https://doi.org/10.5281/ZENODO.5717953}
  {\bibinfo {title} {Data for ``{Comparing} ultrastable lasers at $7\times
  10^{-17}$ fractional frequency instability through a 2220 km optical fibre
  network'' [zenodo.org/record/5717953]}} (\bibinfo {year} {2021})\BibitemShut
  {NoStop}%
\bibitem [{\citenamefont {Hirsch}(2016)}]{Hirsch:2016}%
  \BibitemOpen
  \bibfield  {author} {\bibinfo {author} {\bibfnamefont {M.}~\bibnamefont
  {Hirsch}},\ }\href {https://doi.org/10.5281/zenodo.213676} {\bibinfo {title}
  {Pymap3d:\ {{3D}} coordinate conversions for geospace,
  10.5281/zenodo.213676}} (\bibinfo {year} {2016})\BibitemShut {NoStop}%
\bibitem [{\citenamefont {Roberts}\ \emph {et~al.}(2024)\citenamefont
  {Roberts}, \citenamefont {Caddell},\ and\ \citenamefont
  {Filzinger}}]{GitHubGPS}%
  \BibitemOpen
  \bibfield  {author} {\bibinfo {author} {\bibfnamefont {B.~M.}\ \bibnamefont
  {Roberts}}, \bibinfo {author} {\bibfnamefont {A.~R.}\ \bibnamefont
  {Caddell}},\ and\ \bibinfo {author} {\bibfnamefont {M.}~\bibnamefont
  {Filzinger}},\ }\href {https://github.com/benroberts999/gps-analysis}
  {\bibinfo {title} {{GPS} analysis, github.com/benroberts999/gps-analysis}}
  (\bibinfo {year} {2024})\BibitemShut {NoStop}%
\bibitem [{\citenamefont {VanderPlas}(2018)}]{VanderPlas_2018}%
  \BibitemOpen
  \bibfield  {author} {\bibinfo {author} {\bibfnamefont {J.~T.}\ \bibnamefont
  {VanderPlas}},\ }\bibfield  {title} {\bibinfo {title} {Understanding the
  lomb–scargle periodogram},\ }\href
  {https://doi.org/10.3847/1538-4365/aab766} {\bibfield  {journal} {\bibinfo
  {journal} {The Astrophysical Journal Supplement Series}\ }\textbf {\bibinfo
  {volume} {236}},\ \bibinfo {pages} {16} (\bibinfo {year} {2018})}\BibitemShut
  {NoStop}%
\bibitem [{\citenamefont {Campbell}\ \emph {et~al.}(2021)\citenamefont
  {Campbell}, \citenamefont {McAllister}, \citenamefont {Goryachev},
  \citenamefont {Ivanov},\ and\ \citenamefont {Tobar}}]{Campbell:2020fvq}%
  \BibitemOpen
  \bibfield  {author} {\bibinfo {author} {\bibfnamefont {W.~M.}\ \bibnamefont
  {Campbell}}, \bibinfo {author} {\bibfnamefont {B.~T.}\ \bibnamefont
  {McAllister}}, \bibinfo {author} {\bibfnamefont {M.}~\bibnamefont
  {Goryachev}}, \bibinfo {author} {\bibfnamefont {E.~N.}\ \bibnamefont
  {Ivanov}},\ and\ \bibinfo {author} {\bibfnamefont {M.~E.}\ \bibnamefont
  {Tobar}},\ }\bibfield  {title} {\bibinfo {title} {Searching for {{Scalar Dark
  Matter}} via {{Coupling}} to {{Fundamental Constants}} with {{Photonic}},
  {{Atomic}}, and {{Mechanical Oscillators}}},\ }\href
  {https://doi.org/10.1103/PhysRevLett.126.071301} {\bibfield  {journal}
  {\bibinfo  {journal} {Phys. Rev. Lett.}\ }\textbf {\bibinfo {volume} {126}},\
  \bibinfo {pages} {071301} (\bibinfo {year} {2021})},\ \Eprint
  {https://arxiv.org/abs/2010.08107} {arXiv:2010.08107} \BibitemShut {NoStop}%
\bibitem [{\citenamefont {Centers}\ \emph {et~al.}(2021)\citenamefont
  {Centers}, \citenamefont {Blanchard}, \citenamefont {Conrad}, \citenamefont
  {Figueroa}, \citenamefont {Garcon}, \citenamefont {Gramolin}, \citenamefont
  {Jackson~Kimball}, \citenamefont {Lawson}, \citenamefont {Pelssers},
  \citenamefont {Smiga}, \citenamefont {Sushkov}, \citenamefont {Wickenbrock},
  \citenamefont {Budker},\ and\ \citenamefont {Derevianko}}]{Centers:2019dyn}%
  \BibitemOpen
  \bibfield  {author} {\bibinfo {author} {\bibfnamefont {G.~P.}\ \bibnamefont
  {Centers}}, \bibinfo {author} {\bibfnamefont {J.~W.}\ \bibnamefont
  {Blanchard}}, \bibinfo {author} {\bibfnamefont {J.}~\bibnamefont {Conrad}},
  \bibinfo {author} {\bibfnamefont {N.~L.}\ \bibnamefont {Figueroa}}, \bibinfo
  {author} {\bibfnamefont {A.}~\bibnamefont {Garcon}}, \bibinfo {author}
  {\bibfnamefont {A.~V.}\ \bibnamefont {Gramolin}}, \bibinfo {author}
  {\bibfnamefont {D.~F.}\ \bibnamefont {Jackson~Kimball}}, \bibinfo {author}
  {\bibfnamefont {M.}~\bibnamefont {Lawson}}, \bibinfo {author} {\bibfnamefont
  {B.}~\bibnamefont {Pelssers}}, \bibinfo {author} {\bibfnamefont {J.~A.}\
  \bibnamefont {Smiga}}, \bibinfo {author} {\bibfnamefont {A.~O.}\ \bibnamefont
  {Sushkov}}, \bibinfo {author} {\bibfnamefont {A.}~\bibnamefont
  {Wickenbrock}}, \bibinfo {author} {\bibfnamefont {D.}~\bibnamefont
  {Budker}},\ and\ \bibinfo {author} {\bibfnamefont {A.}~\bibnamefont
  {Derevianko}},\ }\bibfield  {title} {\bibinfo {title} {Stochastic
  fluctuations of bosonic dark matter},\ }\href
  {https://doi.org/10.1038/s41467-021-27632-7} {\bibfield  {journal} {\bibinfo
  {journal} {Nat. Commun.}\ }\textbf {\bibinfo {volume} {12}},\ \bibinfo
  {pages} {7321} (\bibinfo {year} {2021})},\ \Eprint
  {https://arxiv.org/abs/1905.13650} {arXiv:1905.13650} \BibitemShut {NoStop}%
\bibitem [{\citenamefont {Touboul}\ \emph {et~al.}(2017)\citenamefont {Touboul}
  \emph {et~al.}}]{Touboul:2017grn}%
  \BibitemOpen
  \bibfield  {author} {\bibinfo {author} {\bibfnamefont {P.}~\bibnamefont
  {Touboul}} \emph {et~al.},\ }\bibfield  {title} {\bibinfo {title}
  {{{MICROSCOPE Mission}}:\ {{First Results}} of a {{Space Test}} of the
  {{Equivalence Principle}}},\ }\href
  {https://doi.org/10.1103/PhysRevLett.119.231101} {\bibfield  {journal}
  {\bibinfo  {journal} {Phys. Rev. Lett.}\ }\textbf {\bibinfo {volume} {119}},\
  \bibinfo {pages} {231101} (\bibinfo {year} {2017})},\ \Eprint
  {https://arxiv.org/abs/1712.01176} {arXiv:1712.01176} \BibitemShut {NoStop}%
\bibitem [{\citenamefont {Berg{\'e}}\ \emph {et~al.}(2018)\citenamefont
  {Berg{\'e}}, \citenamefont {Brax}, \citenamefont {M{\'e}tris}, \citenamefont
  {{Pernot-Borr{\`a}s}}, \citenamefont {Touboul},\ and\ \citenamefont
  {Uzan}}]{Berge:2017ovy}%
  \BibitemOpen
  \bibfield  {author} {\bibinfo {author} {\bibfnamefont {J.}~\bibnamefont
  {Berg{\'e}}}, \bibinfo {author} {\bibfnamefont {P.}~\bibnamefont {Brax}},
  \bibinfo {author} {\bibfnamefont {G.}~\bibnamefont {M{\'e}tris}}, \bibinfo
  {author} {\bibfnamefont {M.}~\bibnamefont {{Pernot-Borr{\`a}s}}}, \bibinfo
  {author} {\bibfnamefont {P.}~\bibnamefont {Touboul}},\ and\ \bibinfo {author}
  {\bibfnamefont {J.-P.}\ \bibnamefont {Uzan}},\ }\bibfield  {title} {\bibinfo
  {title} {{{MICROSCOPE Mission}}:\ {{First Constraints}} on the {{Violation}}
  of the {{Weak Equivalence Principle}} by a {{Light Scalar Dilaton}}},\ }\href
  {https://doi.org/10.1103/PhysRevLett.120.141101} {\bibfield  {journal}
  {\bibinfo  {journal} {Phys. Rev. Lett.}\ }\textbf {\bibinfo {volume} {120}},\
  \bibinfo {pages} {141101} (\bibinfo {year} {2018})},\ \Eprint
  {https://arxiv.org/abs/1712.00483} {arXiv:1712.00483} \BibitemShut {NoStop}%
\bibitem [{\citenamefont {Derevianko}(2018)}]{Derevianko:2016vpm}%
  \BibitemOpen
  \bibfield  {author} {\bibinfo {author} {\bibfnamefont {A.}~\bibnamefont
  {Derevianko}},\ }\bibfield  {title} {\bibinfo {title} {Detecting dark-matter
  waves with a network of precision-measurement tools},\ }\href
  {https://doi.org/10.1103/PhysRevA.97.042506} {\bibfield  {journal} {\bibinfo
  {journal} {Phys. Rev. A}\ }\textbf {\bibinfo {volume} {97}},\ \bibinfo
  {pages} {042506} (\bibinfo {year} {2018})},\ \Eprint
  {https://arxiv.org/abs/1605.09717} {arXiv:1605.09717} \BibitemShut {NoStop}%
\bibitem [{\citenamefont {Li}\ \emph {et~al.}()\citenamefont {Li},
  \citenamefont {Liu}, \citenamefont {Dailey},\ and\ \citenamefont
  {Afshordi}}]{Li:2023qpx}%
  \BibitemOpen
  \bibfield  {author} {\bibinfo {author} {\bibfnamefont {Y.}~\bibnamefont
  {Li}}, \bibinfo {author} {\bibfnamefont {R.}~\bibnamefont {Liu}}, \bibinfo
  {author} {\bibfnamefont {C.}~\bibnamefont {Dailey}},\ and\ \bibinfo {author}
  {\bibfnamefont {N.}~\bibnamefont {Afshordi}},\ }\bibfield  {title} {\bibinfo
  {title} {Detecting cosmological scalar fields using orbital networks of
  quantum sensors},\ }\href@noop {} {\ }\Eprint
  {https://arxiv.org/abs/2311.17873} {arXiv:2311.17873} \BibitemShut {NoStop}%
\bibitem [{\citenamefont {{D. Murphy {\sl et al.}}}(2015)}]{MurphyJPL2015}%
  \BibitemOpen
  \bibfield  {author} {\bibinfo {author} {\bibnamefont {{D. Murphy {\sl et
  al.}}}},\ }\href
  {http://kb.igs.org/hc/en-us/articles/221690387-Technical-Report-2015}
  {\bibinfo {title} {{{JPL Analysis Center Technical Report}} 2015 (in {{IGS
  Technical Report}} 2015)}} (\bibinfo {year} {2015})\BibitemShut {NoStop}%
\bibitem [{\citenamefont {{Jet Propulsion Laboratory}}()}]{JPLigsac}%
  \BibitemOpen
  \bibfield  {author} {\bibinfo {author} {\bibnamefont {{Jet Propulsion
  Laboratory}}},\ }\href {https://sideshow.jpl.nasa.gov/pub/jpligsac/}
  {\bibinfo {title} {https://sideshow.jpl.nasa.gov/pub/jpligsac/}}\BibitemShut
  {NoStop}%
\bibitem [{\citenamefont {Blewitt}(2015)}]{Blewitt2015307}%
  \BibitemOpen
  \bibfield  {author} {\bibinfo {author} {\bibfnamefont {G.}~\bibnamefont
  {Blewitt}},\ }\bibfield  {title} {\bibinfo {title} {{{GPS}} and space-based
  geodetic methods},\ }in\ \href
  {https://doi.org/10.1016/B978-0-444-53802-4.00060-9} {\emph {\bibinfo
  {booktitle} {Treatise on {{Geophysics}}}}}\ (\bibinfo  {publisher}
  {{Elsevier}},\ \bibinfo {address} {{Oxford}},\ \bibinfo {year} {2015})\ pp.\
  \bibinfo {pages} {307--338}\BibitemShut {NoStop}%
\bibitem [{\citenamefont {Delva}\ \emph
  {et~al.}(2017{\natexlab{b}})\citenamefont {Delva}, \citenamefont {Hees},\
  and\ \citenamefont {Wolf}}]{Delva:2017znr}%
  \BibitemOpen
  \bibfield  {author} {\bibinfo {author} {\bibfnamefont {P.}~\bibnamefont
  {Delva}}, \bibinfo {author} {\bibfnamefont {A.}~\bibnamefont {Hees}},\ and\
  \bibinfo {author} {\bibfnamefont {P.}~\bibnamefont {Wolf}},\ }\bibfield
  {title} {\bibinfo {title} {Clocks in {{Space}} for {{Tests}} of {{Fundamental
  Physics}}},\ }\href {https://doi.org/10.1007/s11214-017-0361-9} {\bibfield
  {journal} {\bibinfo  {journal} {Space Sci. Rev.}\ }\textbf {\bibinfo {volume}
  {212}},\ \bibinfo {pages} {1385} (\bibinfo {year}
  {2017}{\natexlab{b}})}\BibitemShut {NoStop}%
\bibitem [{\citenamefont {{Abou El-Neaj}}\ \emph {et~al.}(2020)\citenamefont
  {{Abou El-Neaj}} \emph {et~al.}}]{AEDGE:2019nxb}%
  \BibitemOpen
  \bibfield  {author} {\bibinfo {author} {\bibfnamefont {Y.}~\bibnamefont
  {{Abou El-Neaj}}} \emph {et~al.},\ }\bibfield  {title} {\bibinfo {title}
  {{{AEDGE}}:\ {{Atomic Experiment}} for {{Dark Matter}} and {{Gravity
  Exploration}} in {{Space}}},\ }\href
  {https://doi.org/10.1140/epjqt/s40507-020-0080-0} {\bibfield  {journal}
  {\bibinfo  {journal} {EPJ Quantum Technol.}\ }\textbf {\bibinfo {volume}
  {7}},\ \bibinfo {pages} {6} (\bibinfo {year} {2020})},\ \Eprint
  {https://arxiv.org/abs/1908.00802} {arXiv:1908.00802} \BibitemShut {NoStop}%
\bibitem [{\citenamefont {Derevianko}\ \emph {et~al.}()\citenamefont
  {Derevianko}, \citenamefont {Gibble}, \citenamefont {Hollberg}, \citenamefont
  {Newbury}, \citenamefont {Oates}, \citenamefont {Safronova}, \citenamefont
  {Sinclair},\ and\ \citenamefont {Yu}}]{Derevianko:2021kye}%
  \BibitemOpen
  \bibfield  {author} {\bibinfo {author} {\bibfnamefont {A.}~\bibnamefont
  {Derevianko}}, \bibinfo {author} {\bibfnamefont {K.}~\bibnamefont {Gibble}},
  \bibinfo {author} {\bibfnamefont {L.}~\bibnamefont {Hollberg}}, \bibinfo
  {author} {\bibfnamefont {N.~R.}\ \bibnamefont {Newbury}}, \bibinfo {author}
  {\bibfnamefont {C.}~\bibnamefont {Oates}}, \bibinfo {author} {\bibfnamefont
  {M.~S.}\ \bibnamefont {Safronova}}, \bibinfo {author} {\bibfnamefont {L.~C.}\
  \bibnamefont {Sinclair}},\ and\ \bibinfo {author} {\bibfnamefont
  {N.}~\bibnamefont {Yu}},\ }\bibfield  {title} {\bibinfo {title} {Fundamental
  {{Physics}} with a {{State-of-the-Art Optical Clock}} in {{Space}}},\ }\href
  {http://arxiv.org/abs/2112.10817} {\ }\Eprint
  {https://arxiv.org/abs/2112.10817} {arXiv:2112.10817} \BibitemShut {NoStop}%
\bibitem [{\citenamefont {Schkolnik}\ \emph {et~al.}(2023)\citenamefont
  {Schkolnik} \emph {et~al.}}]{Schkolnik:2022utn}%
  \BibitemOpen
  \bibfield  {author} {\bibinfo {author} {\bibfnamefont {V.}~\bibnamefont
  {Schkolnik}} \emph {et~al.},\ }\bibfield  {title} {\bibinfo {title} {Optical
  atomic clock aboard an {{Earth-orbiting}} space station ({{OACESS}}):\
  enhancing searches for physics beyond the standard model in space},\ }\href
  {https://doi.org/10.1088/2058-9565/ac9f2b} {\bibfield  {journal} {\bibinfo
  {journal} {Quantum Sci. Technol.}\ }\textbf {\bibinfo {volume} {8}},\
  \bibinfo {pages} {014003} (\bibinfo {year} {2023})},\ \Eprint
  {https://arxiv.org/abs/2204.09611} {arXiv:2204.09611} \BibitemShut {NoStop}%
\bibitem [{\citenamefont {Tsai}\ \emph {et~al.}(2022)\citenamefont {Tsai},
  \citenamefont {Eby},\ and\ \citenamefont {Safronova}}]{Tsai:2021lly}%
  \BibitemOpen
  \bibfield  {author} {\bibinfo {author} {\bibfnamefont {Y.-D.}\ \bibnamefont
  {Tsai}}, \bibinfo {author} {\bibfnamefont {J.}~\bibnamefont {Eby}},\ and\
  \bibinfo {author} {\bibfnamefont {M.~S.}\ \bibnamefont {Safronova}},\
  }\bibfield  {title} {\bibinfo {title} {Direct detection of ultralight dark
  matter bound to the {{Sun}} with space quantum sensors},\ }\href
  {https://doi.org/10.1038/s41550-022-01833-6} {\bibfield  {journal} {\bibinfo
  {journal} {Nat. Astron.}\ }\textbf {\bibinfo {volume} {7}},\ \bibinfo {pages}
  {113} (\bibinfo {year} {2022})},\ \Eprint {https://arxiv.org/abs/2112.07674}
  {arXiv:2112.07674} \BibitemShut {NoStop}%
\bibitem [{\citenamefont {O'Hare}(2020)}]{AxionLimits}%
  \BibitemOpen
  \bibfield  {author} {\bibinfo {author} {\bibfnamefont {C.}~\bibnamefont
  {O'Hare}},\ }\bibfield  {title} {\bibinfo {title} {{AxionLimits
  (\url{https://cajohare.github.io/AxionLimits/}),}}\ }\href
  {https://doi.org/10.5281/zenodo.3932430} {10.5281/zenodo.3932430} (\bibinfo
  {year} {2020})\BibitemShut {NoStop}%
\end{thebibliography}%

\end{document}